\documentclass[trackchanges, twocolumn]{aastex701}
\usepackage{enumitem}
\usepackage{multirow}
\usepackage{booktabs}
\usepackage{CJK}
\usepackage{amsmath}
\usepackage{graphicx}
\usepackage{subcaption}
\usepackage{hyperref}
\usepackage{xspace}

\newcommand{\msun}{\ensuremath{M_\odot}\xspace}

\begin{document}
\begin{CJK*}{UTF8}{gbsn}
\title{Probing the IMF in the Early Universe -- Direct measurements in the Bo{\"o}tes I UFD  with \textit{JWST/NIRCam}}

\author[0000-0001-8470-1725]{Keyi Ding (丁可怿)}
\affiliation{Department of Astronomy, University of Maryland, College Park, MD 20742, USA}
\email[show]{kyding@umd.edu} 

\author[0000-0002-5581-2896]{Mario Gennaro}
\affiliation{Space Telescope Science Institute, 3700 San Martin Drive, Baltimore, MD 21218, USA}
\email{gennaro@stsci.edu}

\author[0000-0001-9364-5577]{Roberto J.\ Avila}
\affiliation{Space Telescope Science Institute, 3700 San Martin Drive, Baltimore, MD 21218, USA}
\email{avila@stsci.edu}

\author[0000-0003-4223-7324]{Massimo Ricotti}
\affiliation{Department of Astronomy, University of Maryland, College Park, MD 20742, USA}
\email{ricotti@umd.edu}

\author[0000-0002-1691-8217]{Rachael L. Beaton}
\affiliation{Space Telescope Science Institute, 3700 San Martin Drive, Baltimore, MD 21218, USA}
\email{rbeaton@stsci.edu}

\author[0000-0003-4850-9589]{Martha L. Boyer}
\affiliation{Space Telescope Science Institute, 3700 San Martin Drive, Baltimore, MD 21218, USA}
\email{mboyer@stsci.edu}

\author[0000-0002-1793-9968]{Thomas M. Brown}
\affiliation{Space Telescope Science Institute, 3700 San Martin Drive, Baltimore, MD 21218, USA}
\email{tbrown@stsci.edu}

\author[0000-0002-0882-7702]{Annalisa Calamida}
\affiliation{Space Telescope Science Institute, 3700 San Martin Drive, Baltimore, MD 21218, USA}
\email{calamida@stsci.edu}

\author[0000-0001-5870-3735]{Santi Cassisi}
\affiliation{INAF – Osservatorio Astronomico di Abruzzo, via M. Maggini snc, I-64100 Teramo, Italy}
\affiliation{INFN – Sezione di Pisa, Universita' di Pisa, Largo Pontecorvo 3, 56127 Pisa, Italy}
\email{santi.cassisi@inaf.it}

\author[0000-0002-0572-8012]{Vedant Chandra}
\affiliation{Center for Astrophysics $\mid$ Harvard \& Smithsonian, 60 Garden St, Cambridge, MA 02138, USA}
\email{vedant.chandra@cfa.harvard.edu}

\author[0000-0002-2970-7435]{Roger E. Cohen}
\affiliation{Eureka Scientific, 2452 Delmer St., Suite 100, Oakland, CA 94602, USA}
\affiliation{Department of Physics and Astronomy, Rutgers the State University of New Jersey, 136 Frelinghuysen Rd., Piscataway, NJ, 08854, USA}
\email{cohenrogere@gmail.com}

\author[0000-0001-6464-3257]{Matteo Correnti}
\affiliation{INAF Osservatorio Astronomico di Roma, Via Frascati 33, 00078, Monteporzio Catone, Rome, Italy}
\affiliation{ASI-Space Science Data Center, Via del Politecnico, I-00133, Rome, Italy}
\email{matteo.correnti@inaf.it}

\author[0000-0002-1763-4128]{Denija Crnojevi\'{c}}
\affiliation{Department of Physics \& Astronomy, University of Tampa, 401 West Kennedy Boulevard, Tampa, FL 33606, USA}
\email{}

\author[0000-0002-6871-1752]{Kareem El-Badry}
\affiliation{Department of Astronomy, California Institute of Technology, 1200 East California Boulevard, Pasadena, CA 91125, USA}
\email{}

\author[0000-0002-7007-9725]{Marla~Geha}
\affiliation{Department of Astronomy, Yale University, New Haven, CT 06520, USA}
\email{marla.geha@yale.edu}

\author[0000-0001-8867-4234]{Puragra Guhathakurta}
\affiliation{Department of Astronomy \& Astrophysics, University of California Santa Cruz, 1156 High Street, Santa Cruz, CA 95064, USA}
\email{raja@ucolick.org}

\author[orcid=0000-0002-3204-1742]{Nitya Kallivayalil}
\affiliation{Department of Astronomy, University of Virginia, 530 McCormick Road, Charlottesville, VA 22904, USA}
\email{njk3r@virginia.edu}

\author[0000-0001-6196-5162]{Evan N. Kirby}
\affiliation{University of Notre Dame, Department of Physics \& Astronomy, 225 Nieuwland Science Hall, Notre Dame, IN 46556}
\email{ekirby@nd.edu}

\author[0000-0001-5538-2614]{Kristen.~B.~W. McQuinn}
\affiliation{Space Telescope Science Institute, 3700 San Martin Drive, Baltimore, MD 21218, USA}
\affiliation{Rutgers University, Department of Physics and Astronomy, 136 Frelinghuysen Road, Piscataway, NJ 08854, USA} 
\email{kmcquinn@stsci.edu}

\author[0000-0002-1445-4877]{Alessandro Savino}
\affiliation{Department of Astronomy, University of California, Berkeley, Berkeley, CA 94720, USA}
\email{asavino@berkeley.edu}

\author[0000-0003-1247-9349]{Cheyanne Shariat}
\affiliation{Department of Astronomy, California Institute of Technology, 1200 East California Boulevard, Pasadena, CA 91125, USA}
\email{cshariat@caltech.edu}

\author[0000-0002-4733-4994]{Joshua D. Simon}
\affiliation{Observatories of the Carnegie Institution for Science, 813 Santa Barbara Street, Pasadena, CA 91101, USA}
\email{jsimon@carnegiescience.edu}

\author[0000-0002-6442-6030]{Daniel R. Weisz}
\affiliation{Department of Astronomy, University of California, Berkeley, Berkeley, CA 94720, USA}
\email{dan.weisz@berkeley.edu}

\begin{abstract}

The dependence of the stellar initial mass function (IMF) on star-formation environment, particularly at low metallicities and high redshifts, remains poorly constrained. Ultra-faint dwarf galaxies (UFDs) are local fossils of high-redshift galaxies hosting old, metal-poor populations, and their resolved stellar populations provide unique pathways to constrain the sub-solar IMF. We investigate the low-mass IMF in the Bo{\"o}tes I (Boo I) UFD with JWST/NIRCam, leveraging its capability to resolve over 10,000 stars reaching $\lesssim$0.15\msun, obtaining one of the largest, deepest resolved stellar samples for UFDs. We explore three different functional forms of the IMF with machine learning and statistical techniques, combining forward modeling of synthetic color-magnitude diagrams with simulation-based inference. We find that a single power-law IMF provides a poorer description of the observed luminosity function and yields a slope inconsistent with the canonical Salpeter IMF. Our best-fit broken power-law and lognormal IMF parameters are consistent with the Milky Way within 68\% confidence level, providing evidence that star formation at metallicities as low as $\mathrm{[Fe/H]}\approx-2.4$ follows a similar IMF as in the Milky Way. By treating Boo I as a local relic analogous to a high-redshift galaxy with a stellar mass of $\lesssim10^5\msun$ at $z\gtrsim6$, our results provide evidence for the universality of the IMF across both local and high-redshift environments.

\end{abstract}

\keywords{\uat{Initial mass function}{796} --- \uat{Stellar astronomy}{1583} --- \uat{Low mass stars}{2050} ---\uat{Dwarf galaxies}{416} ---\uat{Infrared photometry}{792}}


\section{Introduction} 

The stellar initial mass function (IMF) is an empirical distribution that describes the masses of stars at birth in a given environment, yet its dependence on the star formation environment--particularly at low metallicities--remains poorly constrained. As a cornerstone of stellar population studies, it connects to many subfields of astrophysics. For instance, on the high-mass end, the IMF determines the number of core-collapse supernovae, gamma-ray bursts, and black hole mergers, setting the expected rate of these energetic events (e.g., \citealt{wea21}). On the low-mass end, the IMF governs the estimation of galaxies' stellar masses.

Currently, the nature and variation of the low-mass IMF across different star formation environments are subject to debate both observationally and theoretically, and the physics of turbulence, radiative feedback, and fragmentation within molecular clouds that lead to the IMF remains uncertain. Theoretically, the metallicity determines the major cooling mechanisms and the cooling rate of the protostellar molecular cloud: metal-poor environments are primarily cooled by molecular hydrogen, whereas metal-rich environments are dominated by dust cooling. This shift in cooling physics alters the Jeans mass and, consequently, the shape of the IMF (e.g., \citealt{bia19, sha22, kle23}). However, some simulations suggest that such variation is weak and difficult to detect (e.g., \citealt{mye11, bat14, bat19, tan24}), and can be erased by stochastic dynamical effects (see \citealt{off14} for a review, and the references therein).

Observationally, \cite{sal55} described the Milky Way (MW) IMF as a single power law with a canonical slope of $\alpha=-2.35$. These early studies, however, lacked the depth to probe the sub-solar regime, and it has since been observed that the single power-law form diverges toward low masses and does not adequately reproduce the observed MW luminosity function. To address this, low-mass IMF models often incorporate a characteristic break mass, implemented either as a broken power law or a log-normal distribution. Representative examples of these approaches include the IMFs proposed by \cite{kro01} and \cite{cha03}. Given that the stellar lifetimes of the low-mass main-sequence stars are comparable to or much longer than the Hubble time, the low-mass IMF can be directly constrained through star counts of resolved stellar populations. 

Before the launch of the Hubble Space Telescope (HST), such resolved studies were limited to environments within the Milky Way, as more distant systems could not be observed at the necessary depth. HST has since overcome this limitation, allowing extragalactic IMF measurements to be directly compared with those of the MW, and observationally test the theory of IMF's dependence on star formation environments. With HST imaging, \cite{wys02} determined a single power-law IMF for the Ursa Minor (UMi) dwarf spheroidal galaxy, and found its luminosity function to be indistinguishable from that of MW globular clusters, suggesting a potentially invariant IMF across systems with markedly different dark-matter content. Using HST imaging, \cite{kal13} investigated the sub-solar IMF of the Small Magellanic Cloud (SMC) and found a similar single power-law slope of $\alpha = -1.9$. However, no peak mass was detected in the luminosity function, making it unclear whether the functional form of the IMF and its overall shape differ in extragalactic star-forming environments. The debate over the existence of a characteristic break mass in the SMC IMF was further complicated by deeper observations with JWST. \cite{leg25} found that introducing a break mass was unnecessary when probing the outskirts of the SMC, whereas \cite{coh26} reported a break mass of $0.49,\msun$ in a broken power-law fit.

Ultra-faint dwarf (UFD) galaxies, characterized by their old, metal-poor, and dark-matter-dominated nature, have since become popular targets for constraining the low-mass IMF in extreme environments using deep HST imaging. Theories and cosmological simulations indicate that UFDs may have formed before or during the epoch of reionization, acting as ‘fossil’ records that retain the properties of high-redshift galaxies \citep{bov09, bov11}; this theory is further supported by deep imaging observations \citep{bro14}. Additionally, with very high mass-to-light ratios, the two-body relaxation times of UFDs exceed the Hubble time, making these systems collisionless. Therefore, unlike globular clusters of similar total mass, where both the total number of stellar particles as well as their spatial density drive the relaxation time well below the Hubble time, UFDs have negligible mass segregation (e.g., \citealt{bau22}), and therefore their stellar mass distributions are not altered by dynamical effects. \cite{geh13} probed the IMF of Hercules and Leo IV with HST/Advanced Camera for Surveys (ACS) down to $\sim0.5$\msun and found single power-law slopes shallower than the \cite{sal55} IMF. \cite{gen18a} reanalyzed the \cite{geh13} data, and extended the sample with four additional UFDs observed with HST/ACS; the analysis probed slightly lower mass (0.45 Msun), and found a mild trend of shallower IMF slope with lower metallicity. \cite{saf22} analyzed the HST/ACS data of Reticulum II for the purpose of detecting wide binaries, and found that a shallower power-law $1.01 \leq \alpha \leq 1.15$ is necessary to explain the mass function. Nonetheless, neither analysis statistically rejects the MW IMF among all their samples. However, in all these studies, the observations were not deep enough to reach the canonical characteristic mass of the MW IMF at $\sim0.2~$\msun, and thus, it remained ambiguous whether a single power-law could describe the luminosity function when extrapolating to lower masses. To overcome the lower mass limitation, multiple studies \citep{gen18, fil22, fil24} analyzed deeper HST data of four UFDs down to $\sim0.2$\msun. \cite{gen18} used WFC3/IR to study Coma Berenices, reaching very low masses in principle, but contamination by background galaxies limited the effective depth. \cite{fil22} targeted Bo{\"o}tes I (Boo I), a more massive and distant UFD, reaching $\sim0.3~$\msun, while \cite{fil24} focused on Reticulum II, Ursa Major II, Triangulum II, and Segue 1 with similar depth. While these studies inferred shallower single power-law slopes, the best-fit broken power-law and lognormal parameters remain consistent with those of the MW within $\sim3\sigma$, albeit with large uncertainties. The small sample size of resolved stars limits the precision with which the low-mass IMF can be constrained. Meanwhile, simulations of the IMF further suggest that fitting a single power-law to an underlying lognormal IMF naturally produces shallower slopes as the lower-mass limit decreases \citep{elb17}, which may explain the range of single power-law slopes reported in studies probing different mass ranges. These limitations illustrate that precisely constraining the low-mass IMF requires both a sufficiently massive UFD and the resolution to reach the canonical MW Chabrier characteristic mass at $\sim0.2~$\msun.

Boo I is a satellite galaxy of the MW first discovered in the Sloan Digital Sky Survey \citep{Bel06}. Spectroscopic studies indicate that it is a dark matter-dominated system, with a mass-to-light ratio of $165\pm45$ \citep{geh26}, a characteristic of UFDs. It hosts predominantly metal-poor stellar populations with a mean metallicity of [Fe/H]$=-2.4\pm0.05$ \citep{geh26, san26} and exhibits the relatively simple chemical enrichment history\citep{fre16, san26}. Boo I can be considered a “true fossil” of high-redshift galaxies \citep{ric05}, with its star formation largely quenched by reionization \citep{bro14}. The finding that nearly 90\% of its stars formed before its first infall into the Milky Way \citep{roc12, bro14, wei15, dur25} further supports its status as a relic of the high-redshift universe that has experienced little contamination from subsequent star formations. Its proximity to the MW \citep[$d = 65~\mathrm{kpc}$;][]{oka12} and relatively higher luminosity among UFDs ($M_V\sim-6.3$ mag) also make it easier to obtain large samples of resolved stars on the low-mass end with deep imaging. Both \cite{gen18a} and \cite{fil22} used HST/ACS imaging (of different depths) to find the IMF of Boo I to be consistent with that of the MW; however, the uncertainties of the IMF parameters in these studies were jointly driven by observational depth, limited sample size, and foreground/background contamination. To overcome these limitations, we exploit the resolving capability of the JWST Near Infrared Camera (NIRCam) to obtain deep imaging of Boo I reaching $\lesssim$0.15\msun and resolve $\sim$10,000 member stars, enabling a more robust measurement of the IMF. 

The paper is organized as follows: Section~\ref{sec:data} describes our observation strategy, source extraction, and artificial star tests. We detail our method to construct stellar models used in subsequent analysis in Section~\ref{sec:model}, and our method for fitting IMF models to our observations in Section~\ref{sec:method}. Our results of various models are summarized in Section~\ref{sec:results}, and we discuss our results in comparison to previous works in Section~\ref{sec:discussions}. Our conclusions are summarized in Section~\ref{sec:conclusions}.

\section{Data}
\label{sec:data}

\subsection{Observation and Image Calibration}

\begin{figure}[t]
                \centering
                 \includegraphics[width=0.9\linewidth]{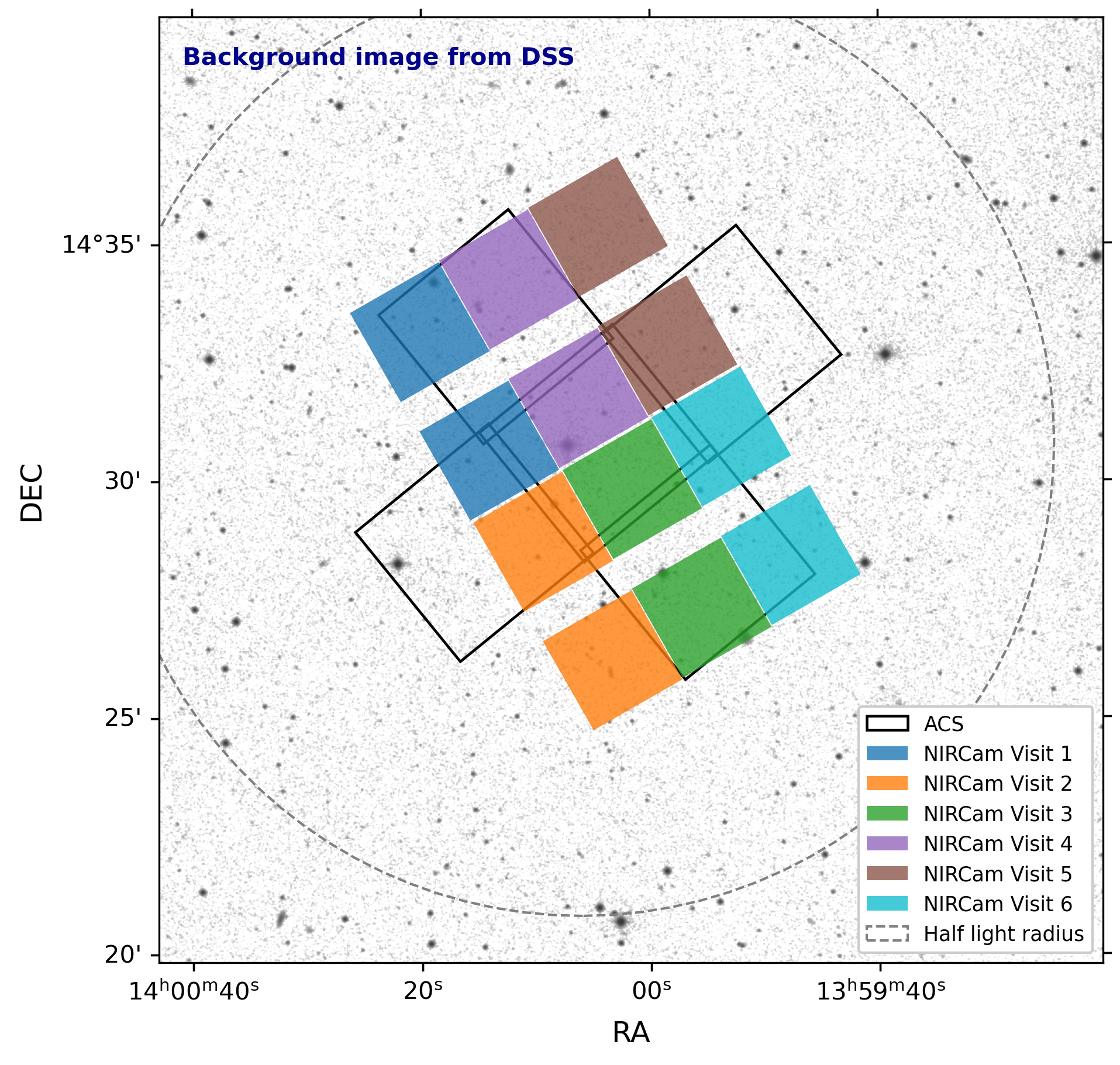}
\caption{Digitized Sky Survey image of Bootes I. The black outlines show the approximate locations of the five ACS visits from \citet{bro14}. The solid colored squares show the locations of our F322W JWST observations. Each color corresponds to a visit, and each pair within a colored pair represents the two NIRCam modules. The dashed circle shows the projected half-light radius based on \cite{mun18}.
\label{fig:footprints}}
\end{figure}

The target fields were selected to overlap with the HST/ACS imaging of Bootes I (GO-12549; PI: Brown) used in \citet{bro14}. Figure~\ref{fig:footprints} shows a Digitized Sky Survey image with the ACS and JWST footprints. We chose filter F150W on the short-wavelength (SW) channel and filter F322W2 on the long-wavelength (LW) channel. These 2 filters offer high sensitivity and provide a color baseline that enables a clear separation of member stars from background galaxies in the color-magnitude diagram (CMD).

The JWST/NIRCam observing setup, including mosaic parameters, dither pattern, and exposure times, is summarized in Table~\ref{tab:obs_setup}. The data were processed using the \texttt{jwst} pipeline version 1.17.1 and calibrated using Calibration Reference Data System (CRDS) \texttt{jwst\_1322.pmap} reference files, with several steps modified from their default behavior. For the SW channel, we enabled the \texttt{clean\_flicker\_noise} step, which removes $1/f$ noise; disabled the \texttt{suppress\_one\_group} step, which retains the zeroth images during the ramp fitting stage; and applied version 3 \citep{sun25} of the Wisp correction algorithm (see the JDOX wisps page\footnote{\url{https://jwst-docs.stsci.edu/known-issues/nircam-known-issues/nircam-scattered-light-artifacts\#NIRCamScatteredLightArtifacts-wispsWisps}}), which removes scattered light artifacts from detectors A3, A4, B3, and B4. For the LW images, the only non-default modification to the pipeline was disabling the \texttt{suppress\_one\_group} step.

\begin{table}[ht]
\centering
\begin{tabular}{ll}
\hline
\hline
\multicolumn{2}{c}{\textbf{Fixed Target}} \\
\hline
Name & BOOI \\
RA (J2000) & 14:00:04.0000 \; ($210.0166667^\circ$) \\
Dec (J2000) & $+14$:30:47.00 \; ($14.51306^\circ$) \\
\hline
\multicolumn{2}{c}{\textbf{Mosaic Parameters}} \\
\hline
Rows & 3 \\
Columns & 2 \\
Row Overlap (\%) & 0.0 \\
Column Overlap (\%) & 0.0 \\
Row Shift (deg) & 0.0 \\
Column Shift (deg) & 0.0 \\
Tile Order & DEFAULT \\
\hline
\multicolumn{2}{c}{\textbf{Dither Pattern}} \\
\hline
Subpixel Dither Type & STANDARD \\
Subpixel Positions & 20 \\
\hline
\multicolumn{2}{c}{\textbf{Spectral Elements}} \\
\hline
Short Filter & F150W \\
Long Filter & F322W2 \\
Readout Pattern & MEDIUM8 \\
Groups/Integration & 9 \\
Integrations/Exposure & 1 \\
Total Integrations & 20 \\
Total Dithers & 20 \\
Total Exposure Time (s) & 18896.715 \\
\hline
\hline
\end{tabular}
\caption{Summary of the JWST observing setup for Boo~I.}
\label{tab:obs_setup}
\end{table}

\subsection{Photometry}
Point-spread function (PSF) fitting photometry was obtained using the DOLPHOT package \citep{dol00, dol16} specific to JWST NIRCam \citep{wei24}. 

Since the F150W images from the NIRCam short-wavelength (SW) channel have a higher spatial resolution and depth than the F322W2 images from the long-wavelength (LW) channel, we broke down the photometry run into 3 steps including the adoption of forced photometry in F322W2 using the warmstart option for optimal source finding: 

\begin{enumerate}[label=(\roman*)]
        \item SW Only: for each SW chip, use the *.i2d image built from all dithers as the reference frame
        \item perform source detection and F150W photometry on the individual *.cal exposures.
        \item SW + LW: feed the source coordinates and signal-to-noise (SNR) from step (ii) into DOLPHOT using the warmstart option. This bypasses source detection, forcing DOLPHOT to perform photometry at the given locations for both the F150W and F322W2 bands.
        \item Cross-match: cross-match the output catalogs from steps (ii) and (iii) by pixel coordinates, retaining the F150W magnitudes from (ii) and the F322W2 magnitudes from (iii).
\end{enumerate}

We find that using a combined *.i2d image from the four SW chips as the reference artificially creates chip-to-chip variations in photometry. We therefore perform photometry on a per-SW-chip basis to mitigate this effect. In our setup, which consists of six non-overlapping mosaic tiles, each with two modules per pointing and four SW chips per module, we divide the photometry run into 48 individual sub-tasks.

The photometry was performed using the VEGA-sirius zero-points\footnote{https://jwst-docs.stsci.edu/jwst-near-infrared-camera/nircam-performance/nircam-absolute-flux-calibration-and-zeropoints}, and the DOLPHOT parameters were determined based on recommendations from the JWST Resolved Stellar Populations Early Release Science Program \citep{wei24}. 

\subsection{Photometric Quality Selection}
\label{sec:qual}
To identify point sources in the DOLPHOT output that are likely stellar members of Boo I, we apply quality cuts to filter out extended sources, suboptimal detections, and contaminations based on a set of DOLPHOT quality parameters, including the Object Type, Flag, Sharpness, Crowding, and SNR.

DOLPHOT classifies each detected source with an Object Type: 1) bright star, 2) faint star, 3) elongated source, 4) object too narrow, or 5) extended source. Additionally, it assigns a Quality Flag for each filter, indicating potential issues: 0 (good measurement), 1 (aperture off-chip), 2 (many bad/saturated pixels), 4 (centrally saturated), or 8 (extreme case). Multiple values can be added to the flag if multiple issues occur. The Sharpness (sharp) measures the angular extent of the source in comparison with the PSF full width at half maximum (FWHM), with a large negative value indicating extended sources such as background galaxies. The Crowding parameter (Crowd) quantifies the level of blending by measuring how much brighter a star would be if neighboring sources were not fit simultaneously.

We use the quality cuts from \cite{war23} and \cite{wei24} as the baseline and add tighter cuts to ensure that the data quality satisfies our main purpose of fitting the IMF. Our final quality criteria include:

\begin{enumerate}
    \item ${\tt Object Type_{SW}}\leq2 \quad {\tt \&}  \quad {\tt Object Type_{LW}}\leq2$
    \item ${\tt Flag_{F150W}}\leq2 \quad {\tt \&}  \quad {\tt Flag_{F322W2}}\leq2$
    \item ${\tt SNR_{F150W}}>12 \quad {\tt \&}  \quad {\tt SNR_{F322W2}}>6$
    \item ${\tt Sharp_{F150W}^2}\leq0.01 \quad {\tt \&}  \quad {\tt Sharp^2_{F322W2}}\leq0.01$
     \item ${\tt Crowd_{F150W}}\leq 0.25 \quad {\tt \&}  \quad {\tt Crowd_{F322W2}}\leq0.5$
     \item we empirically fit a curve of ${\tt \log(\frac{1}{SNR})}$ as a function of magnitude for each band, and reject sources if the difference between the extracted and fitted ${\tt \log(\frac{1}{SNR})}$ is greater than $0.2$ (Appendix~\ref{app:cuts})
\end{enumerate}

 \begin{figure*}[ht]
                \centering
                 \includegraphics[width=0.9\linewidth]{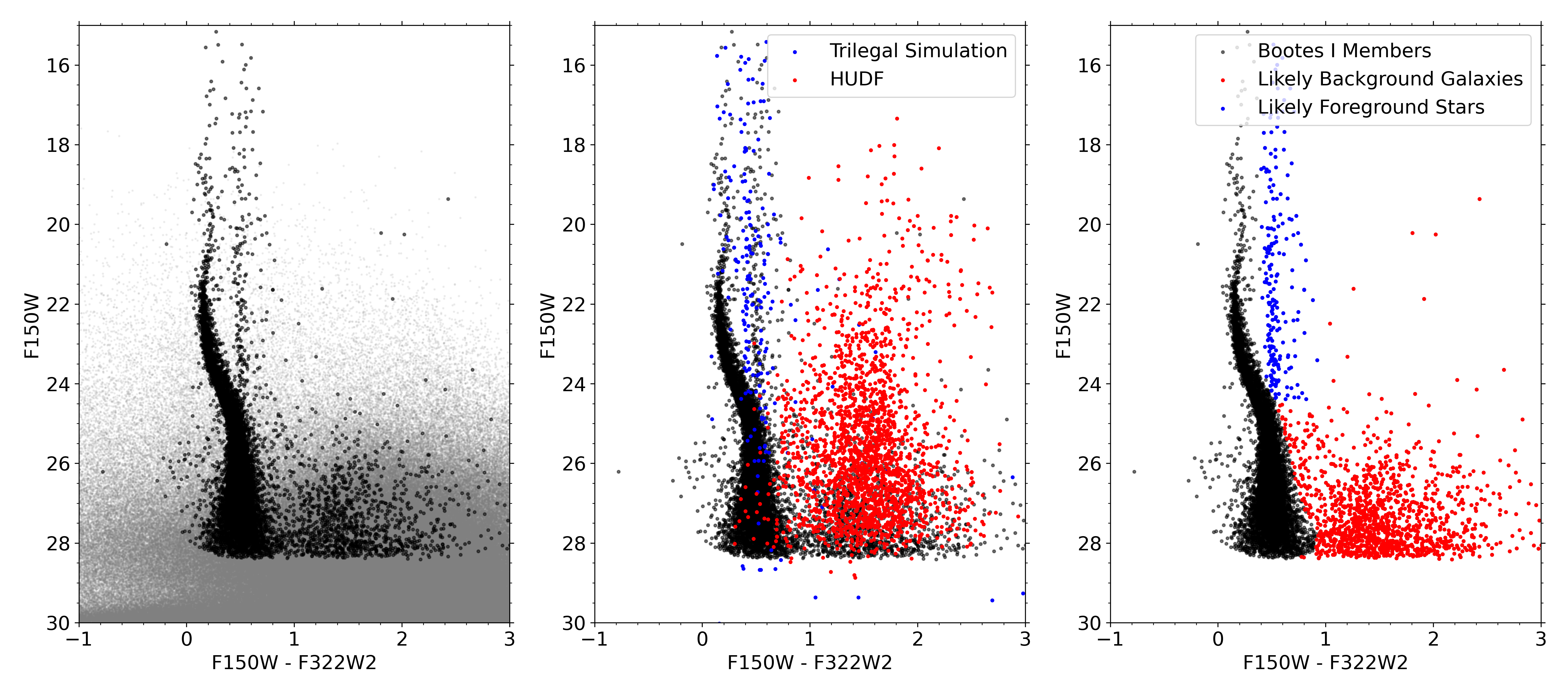}
\caption{CMD of sources extracted by DOLPHOT. Left: All photometric sources are shown in gray, with those passing the data quality cuts highlighted in black. Middle: Sources passing the quality cuts, overplotted with a synthetic foreground stellar population from TRILEGAL (blue) and synthesized photometry of background galaxies from the Hubble Ultra Deep Field (HUDF) based on SED models (red), scaled by area to match the field-of-view of our observations. Right: The final selection of Boo I member stars in black, with likely foreground stars in blue and likely background galaxies in red based on color and magnitude cuts. The Boo I member giants, which we also exclude in our analysis, are plotted in green.
\label{fig:CMD}}
\end{figure*}

The left panel of Figure~\ref{fig:CMD} shows the 12,889 point sources that passed our quality cuts, overlaid on the full sample of 1,376,174 detected sources. This selected sample, however, still contains some contamination from foreground stars and unresolved background galaxies. Specifically, there is a group of unresolved galaxies with $\mathrm{F150W}-\mathrm{F322W2} > 0.8$ and a vertical band of foreground stars at $\mathrm{F150W}-\mathrm{F322W2} \approx 0.6$.

\subsection{Member Stars Identification}
\label{sec:mem}

One advantage of using the SW+LW channels of JWST NIRCam is that they provide a color baseline that substantially reduces contamination from background galaxies, and the F150W+F322W2 filter combination was specifically chosen for this program to maximize color separation. Following a similar method used in \cite{gen18}, we derived the best-fitting spectral template from photometry of HUDF galaxies using the methodology of \cite{pac12, pac13}. These templates, which model galaxy mass, star formation history, and redshift, were then used to calculate synthetic magnitudes in the NIRCam passbands. As the middle panel of Figure~\ref{fig:CMD} shows, HUDF galaxies are almost exclusively redder than the Boo I main sequence, aligning with the contaminated sources in our sample. When the HUDF galaxy count is scaled based on the field-of-view of the HUDF to match our observation, the background contamination is negligible.

To estimate the level of contamination from foreground stars, we use the TRILEGAL model \citep{tri12} to generate a synthetic CMD of the MW stars along the sight of Boo I with an area matching the field-of-view of our observations. The resulting synthetic CMD recovers both the color and density of the vertical band of foreground stars seen in our data. The simulation suggests that at the depth of our observation, the foreground population mostly consists of M-dwarfs and brown dwarfs in the Galactic halo, and they rarely overlap with the main-sequence of Boo I on the CMD.

Our contamination estimate is in good agreement with the prediction shown in Figure~4 of \cite{coh26}, supporting a consistent picture when using the same JWST/NIRCam filters. We apply color and magnitude cuts to the CMD to select likely Boo I member stars. We also impose an upper magnitude limit of $\mathrm{F150W}=21$, as in the turn-off and giant regimes, a narrow mass bin maps to a broad range in magnitude, which can introduce stochastic biases in the inference depending on the sampled magnitudes. The resulting sample of 11,046 candidate members is plotted in black in the right panel of Figure~\ref{fig:CMD}. We divided the candidate member sample among the individual NIRCam detector chips and measured the stellar surface density in each mosaic tile and chip. The density varies by only 3\% (rms) around the mean value, with the largest deviation among individual chips being 4\%. We also find no significant spatial gradient across the field. This demonstrates that the stellar density of the analyzed sample is approximately uniform across the observed footprint, suggesting that spatial variations in crowding are minimal.

\subsection{Chip-dependent Offset Removal}
\label{sec:offset}

 Although per-chip-level calibration significantly reduces chip-to-chip variations, it does not fully eliminate them. We find that these offsets are highly consistent across visits and mosaic tiles, indicating that they are intrinsic to the detector calibration rather than arising from the mosaic construction. To correct for non-negligible photometric offsets among detector chips in JWST/NIRCam, we apply a chip-shift correction to the final catalog after identifying likely Boo I members as described below. For each NIRCam LW module, we divide the stellar catalog into SW chip-specific subsets and select main-sequence stars with $23 < F150W < 28$ to serve as stable fiducial references for measuring chip-to-chip offsets. For each SW chip, we estimate the mean photometric shift relative to a reference chip (a2 for module A and b3 for module B) using bootstrap resampling, performing 5000 iterations with replacement. The resulting shift (ranging from 0.005 to 0.022 magnitude in F150W) is applied to the F150W magnitude, as SW chips sharing the same LW module are not expected to exhibit chip-dependent offsets in F322W2. We then use the same bootstrap procedure to estimate the color offset between the two LW detectors and apply the corresponding correction to the F322W2 magnitude.

\subsection{Artificial Star Tests}
\label{sec:ast}

We conduct artificial star tests (ASTs) to evaluate photometric errors, systematic uncertainties, and completeness. In the ASTs, DOLPHOT injects mock stars with given locations and magnitudes in each band into the images, and attempts to recover the magnitudes along with other photometric parameters using the same setup as the photometry run. We generate the input magnitudes for the artificial star catalog in two steps. First, we sample 10,000 F150W magnitudes between 20 and 30.5 from a power-law luminosity function with a slope of 2. Next, we derive the corresponding colors by numerically interpolating the main-sequence colors of Boo I stars as a function of their F150W magnitude. For each of these 10,000 color-magnitude pairs, we generate 100 realizations by sampling from a Gaussian distribution centered on the interpolated color with a standard deviation of 0.05. This process yields a catalog of 1,000,000 artificial stars. Additionally, we randomly sample another 200,000 stars that uniformly cover the color-magnitude space with $-0.2 < \mathrm{F150W}-\mathrm{F322W2} <1$ and  $16.5 < \mathrm{F150W} < 30.5$. Given the nearly uniform density profile of Boo I within our observed fields as demonstrated in Section~\ref{sec:mem}, we assign spatial coordinates to the artificial stars by uniformly distributing them across the field of view. The magnitude catalog is randomly divided into 48 subsets, one for each SW chip. We process the artificial stars following the same warmstart procedure as the photometry. First, we perform AST on the SW images only. Using the recovered positions from this initial run, we then execute a second AST run on the combined SW and LW images. We apply the same data quality flags and chip-dependent shifting to the artificial stars to identify non-detections and bad measurements.

\begin{figure*}[ht]
                \centering
                 \includegraphics[width=0.6\linewidth]{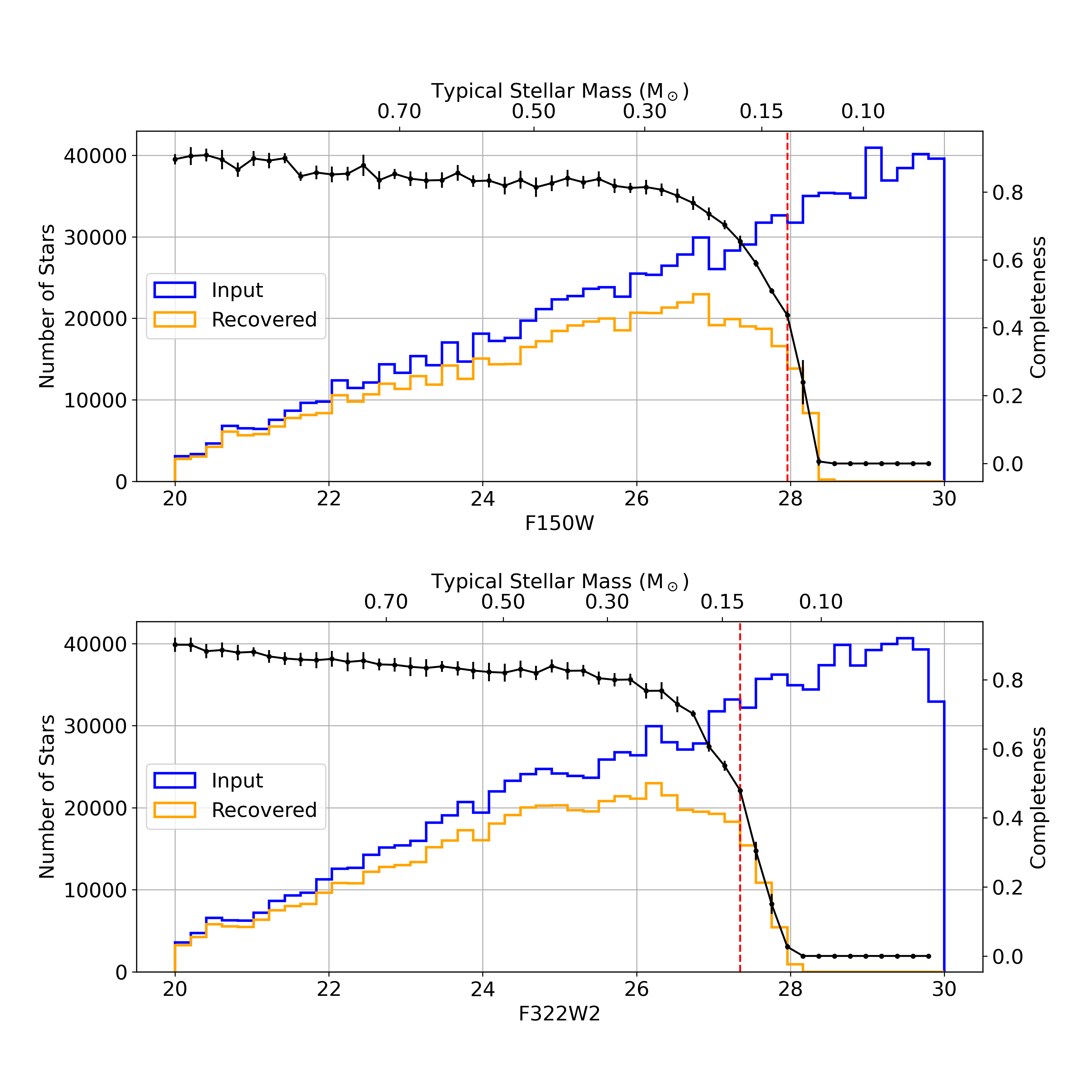}
\caption{Luminosity function of the input magnitudes (blue) and the recovered magnitudes (orange) of the artificial stars in the F150W (top) and F322W2 (bottom) filters. The ratio of these functions (black line) represents the photometric completeness as a function of magnitude, and the error bars represent the rms variation when grouping the ASTs by mosaic tiles and SW chips}. Red vertical dashed lines mark the 50\% completeness limits, while the top x-axis displays the corresponding typical stellar masses based on the stellar model.
\label{fig:AST}
\end{figure*}

Figure~\ref{fig:AST} presents the luminosity functions of the input and recovered stars, along with their completeness ratio. The 50\% completeness level, marked by red dashed lines, occurs at $\mathrm{F150W}=27.9$ and $\mathrm{F322W2}=27.3$. The completeness falls short of 100\% even at the bright end because some artificial stars land on image edges or bad pixels and are therefore not detected. The combined input and output CMDs for the ASTs are shown in Appendix~\ref{app:ast}. To investigate spatial variations in the completeness, we calculated AST completeness separately for each mosaic tile and detector chip. The black curves in Figure~\ref{fig:AST} show the mean completeness as a function of magnitude, with the error bars representing the rms variation when grouping the ASTs by mosaic tiles and SW chips. The average fractional mosaic tile and chip-to-chip variation in completeness is only 4\% over the magnitude range relevant for the IMF analysis, indicating that spatial variations in the selection function are minimal. Moreover, given that Boo I has a sparse spatial distribution, with around 10,000 stars spanning across $2 \times 2040\times2040$ pixels in the LW channel and $8 \times 2040\times2040$ pixels in the LW channel, the completeness is primarily limited by the signal-to-noise instead of crowding. Combined with the nearly uniform stellar surface density across the observed footprint, this demonstrates that spatial variations in the selection function are minimal. Therefore, the use of a global AST kernel is sufficient for our IMF inference.

\section{Stellar Models}
\label{sec:model}

Using our catalog of Boo I member stars, we overlay the CMD with BaSTI isochrones \citep{basti} of various ages, adopting $[\alpha/\mathrm{Fe}] = +0.4$ and $\mathrm{[Fe/H]} = -2.4$, to visually assess the agreement between the data and stellar evolution models. The comparison indicates that the observed CMD exhibits a dimmer turn-off magnitude than predicted by isochrones corresponding to the literature age estimates, and an adequate fit would require an isochrone older than the Hubble time. This discrepancy suggests that a more detailed treatment of chemical abundances is required to reproduce the observed CMD accurately. Observations of metal-poor stars in the MW halo indicate that oxygen abundance increases with decreasing metallicity \citep{fre10}, and high [O/Fe] was also detected in the Segue I UFD \citep{sit21}. The additional oxygen enhancement primarily affects the rate of the CNO cycle, and consequently, variations in the oxygen abundance can modify the main-sequence lifetime, and thus the turnoff age and luminosity. Based on the calculations of \cite{bro14}, increasing [O/Fe] by 0.5 dex would cause the isochrones to appear approximately $\sim1$ Gyr younger when inferred from the turnoff luminosity, which aligns with the discrepancy observed between our data and the BaSTI models.

We use the $\alpha$-enhanced, O-enhanced models used by \cite{bro14}, computed using the Victoria-Regina evolutionary code \citep{van14}. The models have a metallicity range of $-4.0 < \text{[Fe/H]} < -1.0$, an $\alpha$-enhancement of $[\mathrm{\alpha/Fe}] = +0.4$, and an additional oxygen enhancement that increases at low [Fe/H] (see Section 3.2 of \citealt{bro14} for more details). To transfer the models into JWST bands, we calculate synthetic JWST magnitudes from the Model Atmospheres with a Radiative and Convective Scheme \citep[MARCS;][]{marcs} spectra, using the \texttt{synphot} Python package \citep{syn}. While increasing the oxygen abundance can slightly reduce the effective temperature of low-mass stars \citep{van22, van23} and affect their spectral energy distributions \citep{cas20}, these effects diminish with decreasing metallicity and are expected to be smaller than the typical observational uncertainties in the UFD regime. Therefore, we use the $\alpha$-enhanced model grid in the MARCS atmospheres without accounting for the oxygen enhancement in the spectral modeling. It must be noted that, to our knowledge, no synthetic stellar spectral library allows for additional O-enhancement with respect to a baseline $[\mathrm{\alpha/Fe}] = +0.4$,  and thus our choice is not only justified, but in fact a necessity.
 
\begin{figure}[t]
                \centering
                 \includegraphics[width=0.9\linewidth]{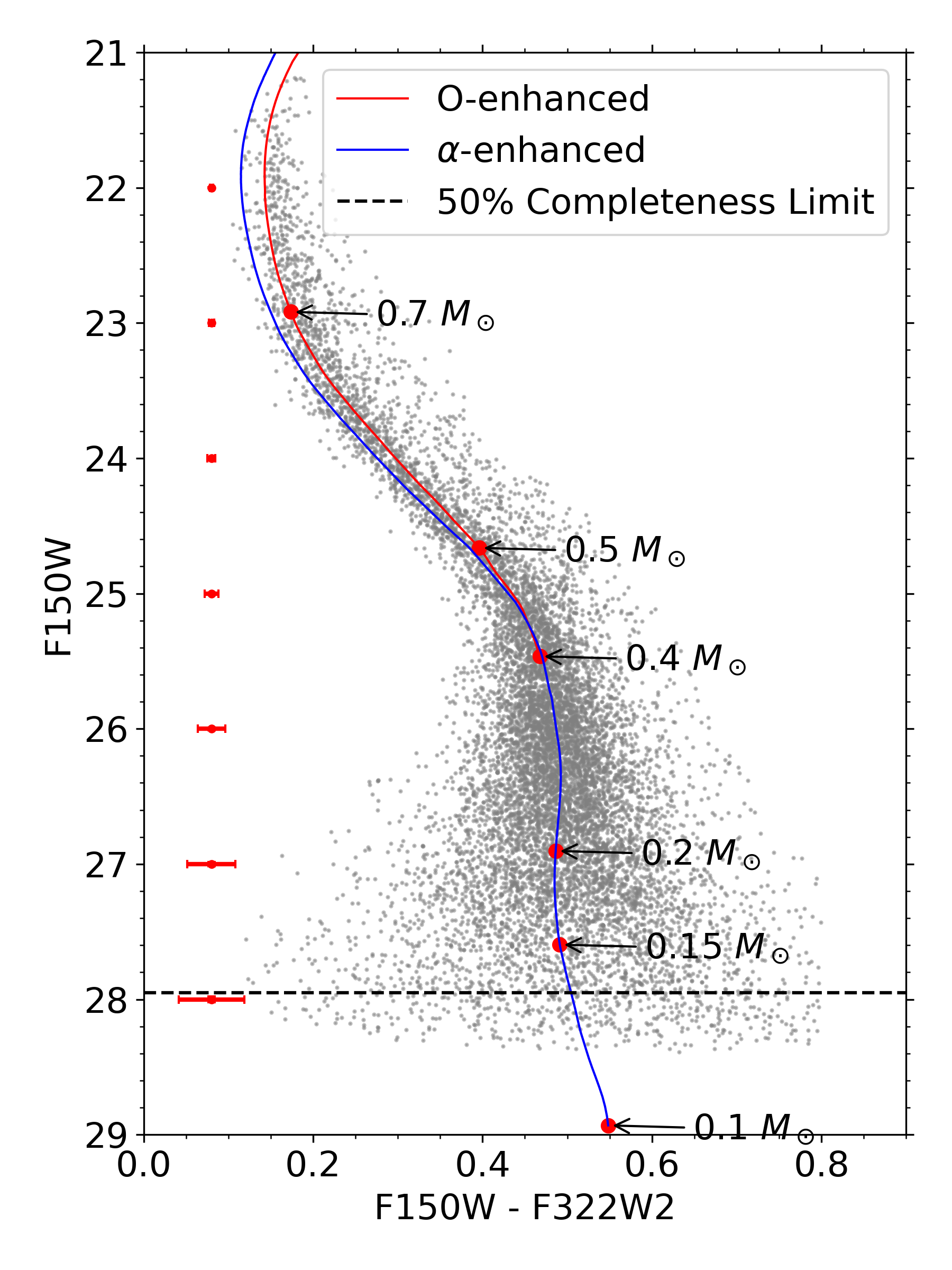}
\caption{CMD of Boo I main-sequence member stars, overlaid with BaSTI $\alpha$-enhanced isochrone (blue) and our O-enhanced isochrone (red), both having an age of 13.3 Gyr and [Fe/H] = -2.4, with labels indicating specific stellar masses. The red error bars indicate the typical photometric uncertainties at each magnitude.
\label{fig:ISO}}
\end{figure}

The oxygen-enhanced model grid is defined only for stellar masses above 0.4\msun. For stars below this limit, we recomputed the models using the $\alpha$-enhanced BaSTI grid \citep{basti}, extending down to 0.1\msun. To create a smooth handoff between the BaSTI and oxygen-enhanced models, we gradually blend their magnitudes over the mass range 0.4\msun to 0.6\msun. The transition is controlled by a cubic Hermite weight function:
 \begin{equation}
    w(m) = \frac{(\mathrm{m_{low}} - \text{m})^2 \left[ 3 - 2 \frac{(\mathrm{m_{low}} - \text{m})}{(\mathrm{m_{low}} - \mathrm{m_{high}})} \right]}{(\mathrm{m_{low}} - \mathrm{m_{high}})^2}
 \end{equation}
where $m_\mathrm{low} = 0.4$\msun, $m_\mathrm{high} = 0.6$\msun, and $w(m)$ is a smooth function ranging from 0 to 1, ensuring that both the value and slope of the interpolated magnitudes match at the boundaries. The interpolated magnitude is then
 \begin{equation}
     \mathrm{mag}(m) = w(m) \times\mathrm{mag_{O-enhanced}} + [1 - w(m)] \times \mathrm{mag_{BaSTI}}.
 \end{equation}
so that stars below $0.4$\msun follow the BaSTI model, those above $0.6$\msun follow the oxygen-enhanced model, and intermediate masses transition smoothly between the two. Figure~\ref{fig:ISO} shows the comparison between the $\alpha$-enhanced O-enhanced isochrones, assuming an extinction $E(B-V)=0.0157$ and a distance of 65 kpc.  We acknowledge that this interpolation introduces some inaccuracy due to differences in the underlying model and the limited calibration of the metal-poor, $\alpha$-enhanced stellar populations, but the resulting magnitude shifts are small compared to the typical photometric uncertainties at the low-mass end.

\section{Method}
\label{sec:method}

We have developed \texttt{Starwave}\footnote{https://github.com/Resolved-Stellar-Populations/starwave} \citep{starwave}, a Python-based code that performs Bayesian inference to determine stellar population parameter sets. We keep the core logic of the method used in \cite{gen18a}: proposing parameters from a prior, simulating corresponding CMDs, and evaluating them against the observed CMD. However, Starwave implements an alternative to the traditional ABC-MCMC method in \cite{gen18a}. Instead, we employ Simulation-based Inference \citep[SBI;][]{cra20}, leveraging neural networks to achieve more efficient parameter estimation. (The initial development of \texttt{Starwave}, built on the \texttt{sbi} Python package \citep{sbi}, was presented at the 237th AAS conference by \citealt{cha21})

\subsection{Simulating Synthetic CMDs}

In the forward modeling process, we use a simulator to generate synthetic CMDs based on a set of stellar populations and IMF parameters. The simulator begins by drawing samples for IMF parameters, binary fraction, star formation history (SFH, includes age and [Fe/H]), distance modulus, and extinction from provided distributions. With the drawn IMF parameters, we sample a set of individual stellar system masses and randomly assign them stellar ages and [Fe/H] based on the drawn SFH. Then we assign absolute magnitudes to each star by interpolating the grid of stellar isochrones, and convert to apparent magnitudes by adding distance modulus and extinction. Under the assumption that most binary stars are unresolved \citep{sha2S}, we model them by calculating their mass ratio, interpolating individual stellar magnitudes, converting magnitudes to fluxes, summing the fluxes, and finally converting back to a combined system magnitude. We assume a constant binary fraction for all system masses and mass ratios. The synthetic CMD is further processed by adding realistic observational noise sampled from the ASTs (which also accounts for completeness) and applying quality cuts based on the upper magnitude limit and color ranges.

We assume Boo I hosts a single old stellar population formed in a short burst around 13.3 Gyr \citep{bro14}, with a heliocentric distance of $d = 65~\mathrm{kpc}$ \citep[corresponding to a distance modulus of $m-M=19.1$;][]{oka12} and extinction $A_V = 0.049$ from the \cite{sch98} dust map. The metallicities are drawn from the Keck/DEIMOS spectroscopic metallicity distribution function \citep{geh26}, peaking at [Fe/H] = -2.4 and held fixed during fitting (See Appendix~\ref{app:mdf} for more details).

\subsection{Empirical Calibrations for Stellar Population Modeling}
To ensure accurate comparison between our observed CMD and synthetic stellar populations, we apply two critical empirical calibrations: (1) a magnitude-dependent color correction to account for systematic offsets between observed CMD and theoretical isochrones, and (2) additional photometric scatter to match the intrinsic uncertainties present in the observational data. With the synthetic CMD simulator, we generate a fiducial synthetic CMD following the literature log-normal IMF parameters from \cite{fil22} and compare it with the observed CMD to correct for the underlying discrepancy between our model and data empirically.  

\subsubsection{Model-Data Color Correction}
\begin{figure*}[ht]
                \centering
                 \includegraphics[width=0.9\linewidth]{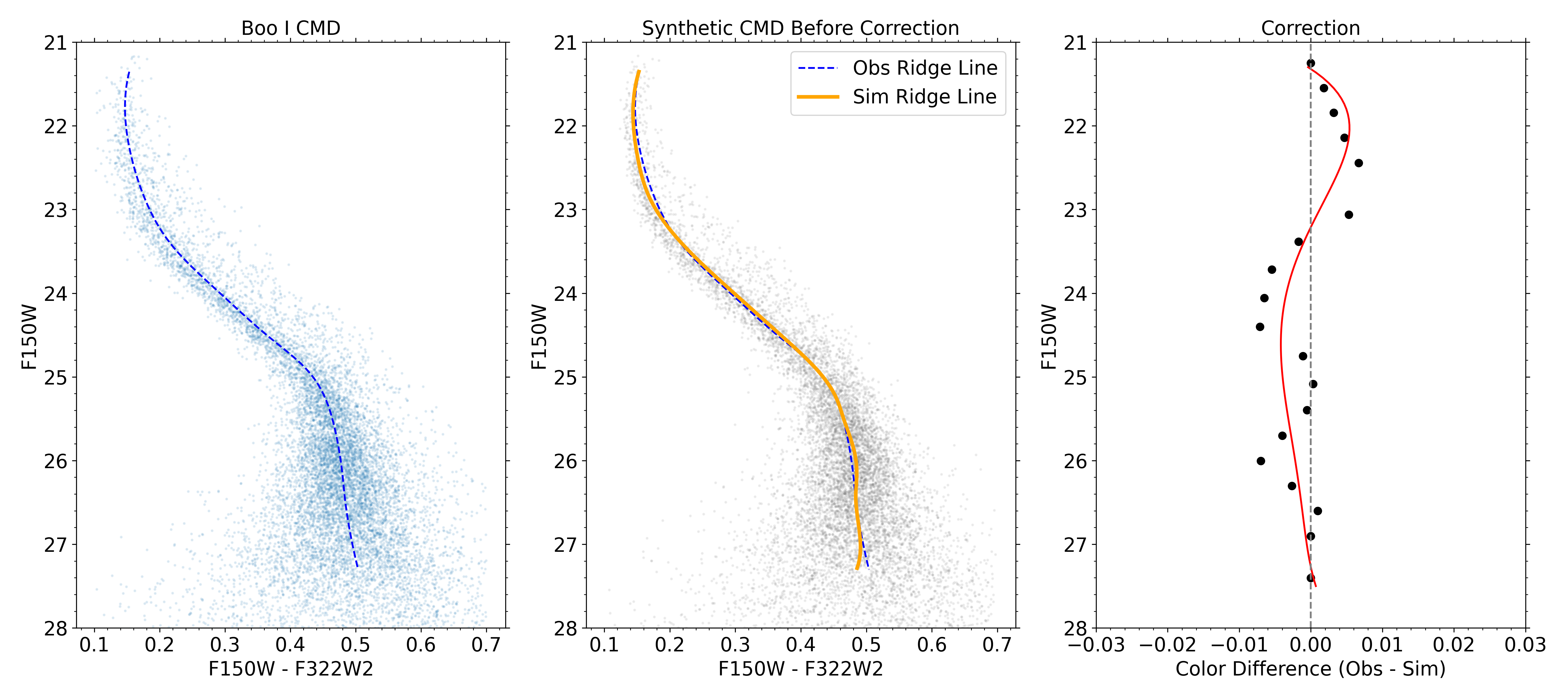}
\caption{Empirical model-data color correction. Left: Observed CMD with ridge line. Middle:
Observed (blue) vs. uncorrected synthetic (red) ridge lines reveal systematic color offset. Right: Magnitude-dependent color correction derived from the ridge line difference.
\label{fig:color_corr}}
\end{figure*}

Despite using the stellar model that accounts for high oxygen abundance as detailed in Section \ref{sec:model}, we still observe a small but non-negligible magnitude-dependent color shift between the fiducial synthetic and observed CMDs. For both the observed and synthetic CMDs, we compute robust ridge lines by dividing the data into magnitude bins with width=0.3 mag spanning the F150W magnitudes from 21 to 28 mag. Within each magnitude bin, we calculate the median color after applying a $3\sigma$ clipping to remove outliers. We then determine the correction at each magnitude bin by applying a Gaussian Kernel Density Estimation to the color distribution with the bandwidth factor determined by the Scott rule \citep{scott}, finding the color with the maximum probability, and computing the difference between the observed and synthetic ridge lines. To construct a smooth, continuous correction function, we fitted a univariate spline to the measured offsets using the \texttt{scipy.interpolate.UnivariateSpline} function. We extended the correction to fainter magnitudes ($27-28.5$ mag) by forcing the offset to approach zero, preventing unphysical extrapolation at the faint limit, as shown in Figure~\ref{fig:color_corr}. This empirical correction typically resulted in shifts ranging from 0 to 0.01 mag across the magnitude range, with the largest corrections applied at the turn-off, where systematic differences are most pronounced. After applying this correction, the synthetic and observed ridge lines agreed to within 0.005 mag across the full magnitude range. Because each magnitude bin used to condition the empirical calibration contains 100 to 1000 stars, the uncertainty on the ridge location, $\sigma_{\text{ridge}} = \sigma_{\text{obs}}/\sqrt{N}$, is at most 10\% of the data uncertainty. Furthermore, to test the sensitivity of the correction to the fiducial IMF used to derive the reference CMD, we draw 500 realizations of the IMF parameters from the approximated posterior distribution of \cite{fil22} and independently calculate the correction function for each realization. We then use the reference points shown in Figure~\ref{fig:ISO} to estimate the uncertainty in the correction at each reference point. When these correction uncertainties are propagated into the photometric color uncertainties, the resulting uncertainties increase by 2.7\% on average and at most 8\%. This reflects the fact that the CMD ridge-line location is primarily determined by the age and metallicity of the stellar population, while the IMF mainly affects the number of stars distributed along the ridge line rather than its location. These correspond to a relatively small contribution compared with the other sources of photometric uncertainty and therefore have a negligible impact on the inferred IMF parameters.

\subsubsection{Additional Photometric Scatter}

The photometric uncertainties in our ASTs do not fully capture the observational uncertainties, as the artificial stars are produced assuming an idealized PSF. To ensure our synthetic populations accurately reproduce the full scatter present in the data, we add magnitude-dependent extra noise to the AST.

We quantify the excess scatter by measuring the width of the color distribution perpendicular to the ridge line for both observed and corrected synthetic CMDs, assuming that the synthetic CMDs captured the physical spread in color from spreads in metallicities, ages, and binary fractions. For magnitude bins of width 0.5 mag from 21 to 28 mag, we compute the perpendicular color residual by subtracting the ridge-line color at each star's magnitude. To account for the asymmetric nature of the color distribution (particularly the right side affected by unresolved binaries), we measure the color dispersion with only the left-side scatter as the difference between the 50th and 16th percentiles. The extra scatter required to match the observations is computed in quadrature:
\begin{equation*}
    \sigma_{\rm extra}(m) = \sqrt{\max(0, \sigma_{\rm obs}^2(m) - \sigma_{\rm sim}^2(m))}
\end{equation*}
where $\sigma_{\rm obs}$ and $\sigma_{\rm sim}$ are the observed and synthetic color dispersions, respectively, with the maximum constraint preventing negative variance. We fit a smooth univariate spline to $\sigma_{\rm extra}(m)$ to obtain a continuous function of magnitude.

To apply this extra scatter to the ASTs catalog while preserving correlated errors between bands, we used the correlation coefficient $\rho(m)$ between photometric errors in F150W and F322W2, measured directly from the original ASTs data in each magnitude bin. For each star in magnitude bin $i$, we added correlated Gaussian noise drawn from a bivariate normal distribution:
\begin{equation*}
\begin{pmatrix} \delta m_{\rm F150W} \\ \delta m_{\rm F322W2} \end{pmatrix} \sim \mathcal{N}\left(\begin{pmatrix} 0 \\ 0 \end{pmatrix}, \begin{pmatrix} \sigma_c^2 & \rho \sigma_c^2 \\ \rho \sigma_c^2 & \sigma_c^2 \end{pmatrix}\right)
\end{equation*}

\noindent
where the per-band noise amplitude is:

\begin{equation}
    \sigma_c = \sigma_{\rm extra} / \sqrt{2(1 - \rho)}
\end{equation}

The measured extra scatter ranged from approximately 0.015 mag at bright magnitudes to 0.035 mag at fainter magnitudes, with correlation coefficients $\rho$ typically between 0.3 and 0.6. After applying both the empirical color correction and the extra scatter, our synthetic CMDs accurately reproduced both the shape and width of the observed CMD across the full magnitude range. These calibrations are applied consistently to all synthetic stellar populations used in our inference framework, ensuring robust constraints on the stellar mass function and other population parameters.

\subsection{Simulation-based Inference}

We use SBI Sequential Neural Posterior Estimation \citep[SNPE;][]{pap16, lue17, gre19} to perform Bayesian parameter estimation for our model, where the likelihood is intractable. This approach trains a neural network to directly approximate the posterior distribution by learning from model simulations, enabling efficient inference from observed data. Here we present a high-level overview of the method:

\begin{enumerate}[label=(\roman*)]
\item Initial Simulation Round: Draw parameter vectors $\theta_i \sim p(\theta)$ from the prior distribution and generate corresponding simulated data $x_i \sim p(x|\theta_i)$ using the forward model. This produces an initial training dataset $\mathcal{D}0 = {(\theta_i, x_i)}$. In our context, $\theta$ denotes the stellar population parameters (e.g., IMF parameters, age, or metallicity), and $x$ corresponds to the synthetic CMD generated by the population synthesis model.
\item Density Estimation: Train a conditional neural density estimator $q_\phi(\theta|x)$ to approximate the posterior distribution $p(\theta|x)$ given samples from $\mathcal{D}0$. The estimator is optimized to minimize the negative log-likelihood $\mathbb{L}(\theta, x) \sim -\log q_\phi(\theta|x)$.
\item Sequential Refinement: Condition the trained model on the observed data $x_{\text{obs}}$ to obtain an approximate posterior $q_\phi(\theta|x_{\text{obs}})$. Use this distribution as the proposal distribution $p_{\text{prop}}(\theta)$ for subsequent simulations, generating new samples $\theta_i \sim p_{\text{prop}}(\theta)$ and $x_i \sim p(x|\theta_i)$. The newly simulated dataset $\mathcal{D}t$ is then combined with or replaces previous rounds to retrain $q_\phi(\theta|x)$, progressively refining the posterior in regions of high posterior density.
\item Inference: After several rounds, the final density estimator $q_\phi(\theta|x_{\text{obs}})$ provides a tractable and differentiable approximation to the true posterior $p(\theta|x_{\text{obs}})$, enabling efficient posterior sampling and uncertainty quantification without further simulations.
\end{enumerate}

\subsection{Kernel Approximation}
\label{sec:kernel}
The SNPE algorithm requires a fixed-length, numerical summary of the observable that is permutation-invariant and captures the data distribution; however, in our case, the number of stars in a CMD is not fixed and the stellar entries are unordered. Therefore, we need a kernel approximation that projects the CMD data points into a finite-dimensional feature vector.

We first Min-Max scale each CMD so that color and magnitude contribute comparably. In the inferences, the scaling limits are determined from the observed CMD and held fixed for subsequent sampling. Then each star $x$ is mapped into the reproducing kernel Hilbert space $\mathcal{H}$ associated with the radial basis function (RBF) kernel:
\begin{equation}
   k(x,y) = \exp(-\gamma|x-y|^2)
\end{equation}
The $\gamma$ controls the kernel bandwidth (equivalently $\gamma$ = 1/(2$\sigma^2$) in which $\sigma$ is the Gaussian width). The RBF is a smooth, shift-invariant similarity: nearby points map to values $\approx$ 1 and distant points map to values $\approx$ 0, so it captures local clustering and continuous structure in CMD space. The choice of $\gamma$ determines the scale of “nearness”: small $\gamma$ (large $\sigma$) produces a very broad kernel that treats many points as similar, emphasizing global structure but smoothing out small-scale features, such as the equal-mass binary sequence; on the other hand, large $\gamma$ (small $\sigma$) makes the kernel very local and sensitive to fine, small-scale differences.

While the kernel is defined on individual stars, our goal is to compare entire CMDs (i.e., sets of stars). We therefore map each catalog to a single representation using a kernel aggregated embedding, defined as the sum of the feature maps of its constituent stars. The RBF kernel corresponds to an implicit mapping into a reproducing kernel Hilbert space $\mathcal{H}$, with $k(x, y)=\langle \phi(x), \phi(y) \rangle_{\mathcal{H}}$. Under this construction, a CMD $\{x_i\}$ is represented by $\mu = \sum_i \phi(x_i)$, so that distances between CMDs A and B correspond to RKHS distances between their embeddings in the Hilbert Space:
\begin{equation}
    D (A, B)=\left\|
        \mu(A)
        -
        \mu(B)
    \right\|_\mathcal{H}
\end{equation}

The feature map $\phi(x)$ is infinite-dimensional, making direct computation infeasible. We therefore approximate it using the Nystr\"{o}m method \citep{wil00}, which selects a set of $n$ landmark points and constructs a low-rank feature map $\phi_{\mathrm{Nys}}(x)\in \mathbb{R}^{n}$ such that
\[
k(x,y) \approx \phi_{\mathrm{Nys}}(x)^\top \phi_{\mathrm{Nys}}(y).
\]
This does not explicitly compute the true infinite-dimensional mapping; instead, it provides a finite-dimensional representation whose inner products approximate the kernel. The per-star feature vectors $\phi_{\mathrm{Nys}}(x_i)$ are then aggregated across the catalog to form a permutation-invariant embedding $\mu_{\mathrm{Nys}}=\sum_i \phi_{\mathrm{Nys}}(x_i) \in \mathbb{R}^n$. In this approximation, distances between CMDs reduce to standard Euclidean distances in $\mathbb{R}^n$, $ D (A, B)=\left\|
        \mu(A)
        -
        \mu(B)
    \right\|_\mathcal{H} 
    \approx\left\|
        \mu_{\mathrm{Nys}}(A)
        -
        \mu_{\mathrm{Nys}}(B)
    \right\|_2$, providing an efficient surrogate for the full kernel-based comparison.

This Nystr\"{o}m+RBF pipeline therefore provides a compact, nonlinear, permutation-invariant summary of CMDs that captures both their global distribution and local geometric structure at scales set by $\gamma$, while remaining computationally efficient. The key hyperparameters are $\gamma$ (kernel width) and $n$ (number of landmark points), which jointly control the trade-off between sensitivity to fine CMD features, approximation accuracy, and runtime. To determine the optimal hyperparameters and quantify the intrinsic uncertainty of our method, we perform convergence tests that recover known input parameters (Appendix~\ref{app:param}).

\subsection{Parameters and Priors}

Using the SBI framework, \texttt{Starwave} simultaneously fits high-dimensional stellar population parameters by sampling from their joint distributions. We fit three different IMF parameterizations, including a single power law (SPL),
\begin{equation}
    p_{\mathrm{SPL}}(m|\alpha) \propto m^\alpha
\end{equation}
a broken power law (BPL),
\[
   p_{\mathrm{BPL}}(m|\alpha_{\mathrm{high}}, \alpha_{\mathrm{low}}, M_b) \propto
  \begin{cases}
    m^{\alpha_{\mathrm{high}}} & m>M_b\\
    km^{\alpha_{\mathrm{low}}} & m<M_b\\
  \end{cases}
\]
\begin{equation}
    k=M_b^{\alpha_{\mathrm{high}}-\alpha_{\mathrm{low}}}
\end{equation}
and a log-normal function (LN)
\begin{equation}
     p_{\mathrm{ln{}}}(m|M_c, \sigma) \propto \frac{1}{m}e^{-\frac{1}{2}(\frac{\log m - log (M_c)}{\sigma})^2}
\end{equation}
In our analysis, we measure the system IMF, in which unresolved multi-star systems are represented as single systems characterized by their total mass. We only consider single and multiple stellar systems and neglect higher-order multiples.

The IMF parameters are also degenerate with the number of stars and the binary fraction, which we include as model parameters. We model the total number of stars as a Poisson distribution with intensity $\log N_{tot}$, and treat $\log N_{tot}$ as a free parameter with a uniform prior from 2 to 6. The binary fraction f(bin) has a uniform prior from 0 to 1, corresponding to the range from no binary stars to all binary stars. By combining these parameters, our model samples their joint distribution, encompassing both the stellar population (\texttt{[$\log N_{tot}$, f(bin)]}) and the IMF parameters. In Appendix~\ref{app:phy}, we also explore the degeneracy of the IMF parameters with the physical stellar population parameters.

\section{Results}
\label{sec:results}

With the photometry, artificial star tests, and the stellar models, we use \texttt{Starwave} to conduct the Bayesian inference of the IMF parameters. We run the inference for 5 rounds with 3000 simulations per round, and we sample the joint distribution of the parameters from the last round as our final posterior.

\begin{figure*}[ht]
\centering

\begin{subfigure}{0.49\textwidth}
    \centering
    \includegraphics[width=\linewidth]{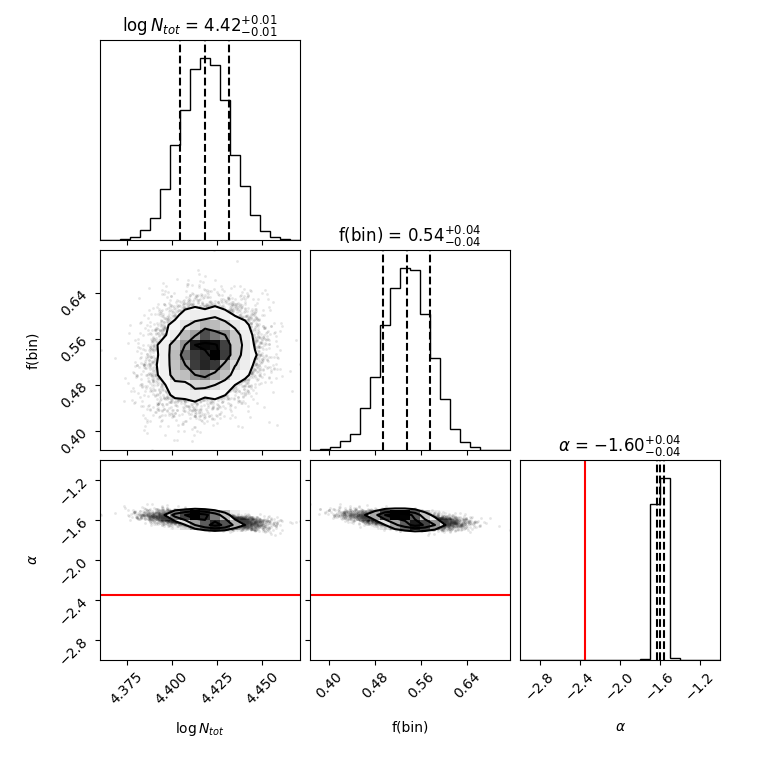}
    \caption{SPL model. The reference slope $\alpha=-2.35$ \citep{sal55} is shown in red.}
    \label{fig:spl_corner}
\end{subfigure}\hfill
\begin{subfigure}{0.49\textwidth}
    \centering
    \includegraphics[width=\linewidth]{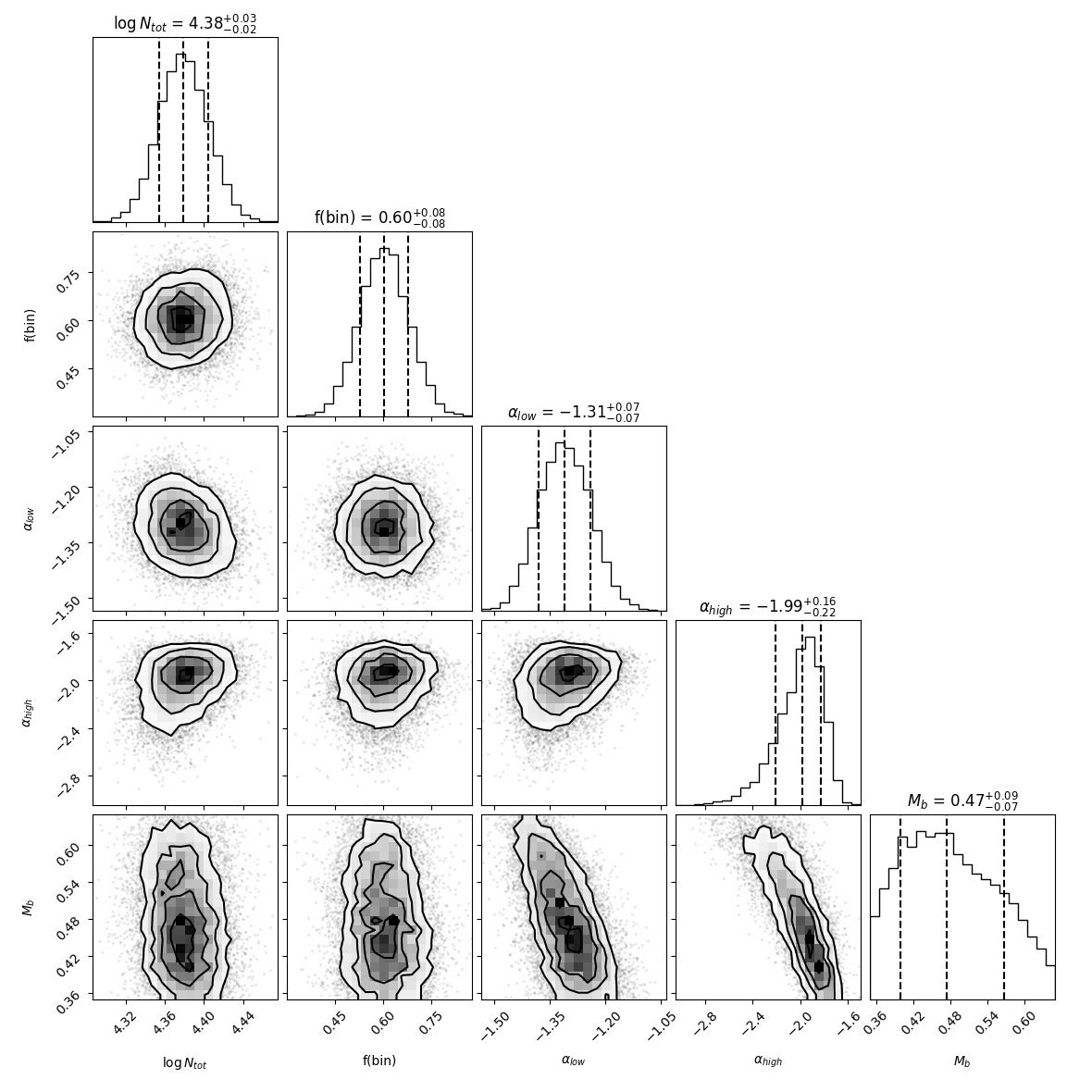}
    \caption{BPL model. \citep{kro01} derived individual-stellar BPL IMF with parameters $\alpha_{\text{low}}=-1.3$, $\alpha_{\text{high}}=-2.3$, $M_b=0.5$\msun. Since our analysis measures the system IMF without resolving or correcting for stellar multiplicity, these parameters are not directly comparable.}
    \label{fig:bpl_corner}
\end{subfigure}

\vspace{0.4cm}

\begin{subfigure}{0.5\textwidth}
    \centering
    \includegraphics[width=\linewidth]{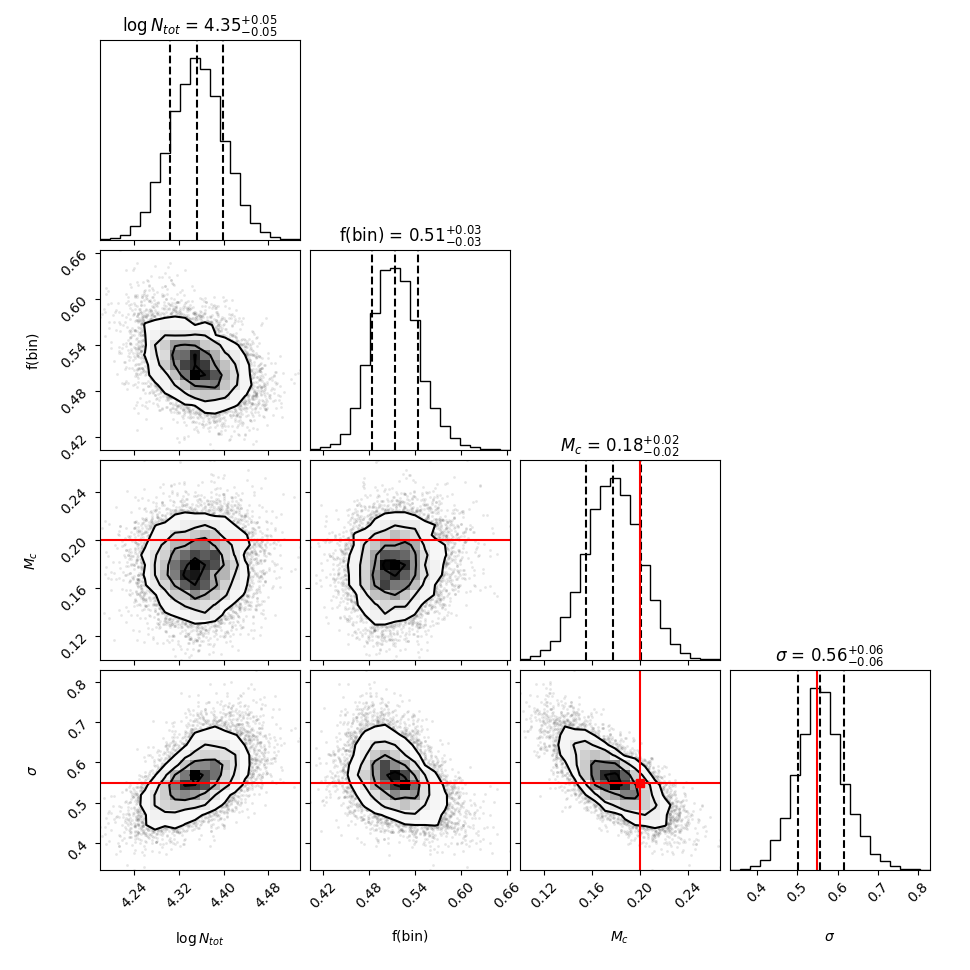}
    \caption{LN model. Reference values $M_c=0.2$\msun, $\sigma=0.55$ \citep{cha03} are shown in red.}
    \label{fig:ln_corner}
\end{subfigure}

\caption{
Corner plots of the posterior distributions for the three IMF models, and the conditional posterior of the BPL model with $M_b \in 0.5 \pm 0.05$. Vertical dashed lines indicate the 16th, 50th, and 84th percentiles of the marginal distributions. 
Red lines denote reference measurements from the solar neighborhood if available.
}
\label{fig:imf_corner_all}

\end{figure*}

The joint distribution of the best-fit parameters as shown in Figure~\ref{fig:spl_corner}, \ref{fig:bpl_corner}, \ref{fig:ln_corner} for the SPL, BPL, LN models, respectively. In each plot, the red line indicates the canonical reference value of the IMF parameters from the solar neighborhood \citep{sal55, kro01,cha03}. In Table \ref{tab:posterior}, we present the median as well as the 68\%, 95\%, and 99\% confidence intervals (CI) of the marginal distributions of the posteriors.

\subsection{Comparison to the Canonical MW Models}

We find our best-fit single power-law slope to be $\alpha=-1.60^{+0.04}_{-0.04}$, which is inconsistent with the \cite{sal55} slope of $\alpha=-2.35$ at a 99\% confidence level. However, we note that a single power-law model is not a valid description of the very low mass IMF due to its diverging nature, and the canonical Salpeter IMF was derived from a sample with very limited depth. As analyzed in \cite{elb17}, if the true underlying IMF follows a log-normal, fitting an SPL over an extended low-mass range leads to an artificially shallower inferred slope as the observation gets deeper. Since our analysis reaches well below the characteristic mass or inflection point identified by \cite{kro01} (and the flattening mass in \citealt{cha03}), it is therefore unsurprising that the SPL model does not to capture the IMF shape and yields a slope inconsistent with \cite{sal55}. A similar trend is evident in our later comparison with results from other literature analyses of varying depth.

Our analysis constrains the break mass and the two power-law slopes under the assumption that a break exists within the range 0.35-0.65 \msun. Our best-fit broken power-law parameters are $\alpha_{low}=-1.31^{+0.07}_{-0.07}$ below the break mass, $\alpha_{high}=-1.99^{+0.16}_{-0.22}$ above the break mass, and the break point following $M_b=0.47^{+0.09}_{-0.07} \mathrm{M_\odot}$. We apply a relatively narrow range to the prior for the break mass, from 0.35\msun to 0.65\msun, because our data span only a small mass range and provide only weak constraints on the break mass. Within this restricted break-mass range, we compare the BPL and SPL models to assess whether the BPL provides a better description of the data. This comparison is conditional on the presence of a break within the adopted mass range and therefore does not by itself establish that the data require a break. Allowing the break mass to vary over a broader range, comparable to or larger than the observed mass range, would cause the sampler to push it toward the boundaries of the data if the sampler and machine learning model exhibit potential underlying bias or overfitting. In that case, the model effectively reduces to a single power law, leaving the second slope poorly constrained. This degeneracy is demonstrated through a convergence test in which we generate a synthetic CMD with known input parameters and use our pipeline to recover those parameters while allowing the break mass to vary over a broad prior range. The details of this test are presented in Appendix~\ref{app:bm}.

\citet{kro01} derived a broken power-law IMF for individual stellar masses with parameters $\alpha_{\text{low}}=-1.3$, $\alpha_{\text{high}}=-2.3$, $M_b=0.5$\msun. In contrast, our analysis fits the system IMF, in which unresolved multiple-star systems are treated as single systems. Because the mapping between the individual-star and system IMFs depends on the binary fraction and other properties of the binary population, these parameters cannot be compared directly.

The best-fitting log-normal parameters are a characteristic mass $M_c=0.18^{+0.02}_{-0.02}$\msun and a width $\sigma=0.56^{-0.06}_{-0.06}$. As the relative uncertainties reach $\sim10\%$, the result provides strong evidence that the Boo I lognormal IMF is consistent with the \cite{cha03} system IMF values of $ M_c=0.2$\msun and $\sigma=0.55$ within the 68\% confidence interval.

\begin{table}[h]
\centering
\begin{tabular}{|l|c|c|c|c|}
\hline
\multicolumn{5}{|c|}{\textbf{Single Power Law}} \\
\hline
Parameter & Median & 68\% CI & 95\% CI & 99\% CI \\
\hline
$\log N_{tot}$ & $4.42$ & $+0.01$ & $+0.03$ & $+0.04$ \\
 & & $-0.01$ & $-0.03$ & $-0.04$ \\
f(bin) & $0.54$ & $+0.04$ & $+0.08$ & $+0.12$ \\
 & & $-0.04$ & $-0.08$ & $-0.13$ \\
$\alpha$ & $-1.60$ & $+0.04$ & $+0.07$ & $+0.11$ \\
 & & $-0.04$ & $-0.07$ & $-0.11$ \\
\hline
\multicolumn{5}{|c|}{\textbf{Broken Power Law}} \\
\hline
Parameter & Median & 68\% CI & 95\% CI & 99\% CI \\
\hline
$\log N_{tot}$ & $4.38$ & $+0.03$ & $+0.05$ & $+0.08$ \\
 & & $-0.02$ & $-0.05$ & $-0.08$ \\
f(bin) & $0.60$ & $+0.08$ & $+0.15$ & $+0.23$ \\
 & & $-0.08$ & $-0.15$ & $-0.24$ \\
$\alpha_{low}$ & $-1.31$ & $+0.07$ & $+0.14$ & $+0.21$ \\
 & & $-0.07$ & $-0.13$ & $-0.19$ \\
$\alpha_{high}$ & $-1.99$ & $+0.16$ & $+0.27$ & $+0.38$ \\
 & & $-0.22$ & $-0.50$ & $-0.77$ \\
$M_{b}$ & $0.47$ & $+0.09$ & $+0.15$ & $+0.17$ \\
 & & $-0.07$ & $-0.11$ & $-0.12$ \\
\hline
\multicolumn{5}{|c|}{\textbf{Log Normal}} \\
\hline
Parameter & Median & 68\% CI & 95\% CI & 99\% CI \\
\hline
$\log N_{tot}$ & $4.35$ & $+0.05$ & $+0.10$ & $+0.15$ \\
 & & $-0.05$ & $-0.10$ & $-0.15$ \\
f(bin) & $0.51$ & $+0.03$ & $+0.07$ & $+0.12$ \\
 & & $-0.03$ & $-0.06$ & $-0.09$ \\
$M_c$ & $0.18$ & $+0.02$ & $+0.05$ & $+0.07$ \\
 & & $-0.02$ & $-0.05$ & $-0.07$ \\
$\sigma$ & $0.56$ & $+0.06$ & $+0.13$ & $+0.21$ \\
 & & $-0.06$ & $-0.11$ & $-0.17$ \\
\hline
\end{tabular}
\caption{Posterior parameter estimates for all three IMF models, together with the 68\%, 95\%, and 99\% confidence interval (CI) of the marginal posterior distributions.}
\label{tab:posterior}
\end{table}

\subsection{Goodness-of-Fit}
\subsubsection{Qualitative Assessment}

\begin{figure*}[ht]
                \centering
                 \includegraphics[width=0.9\linewidth]{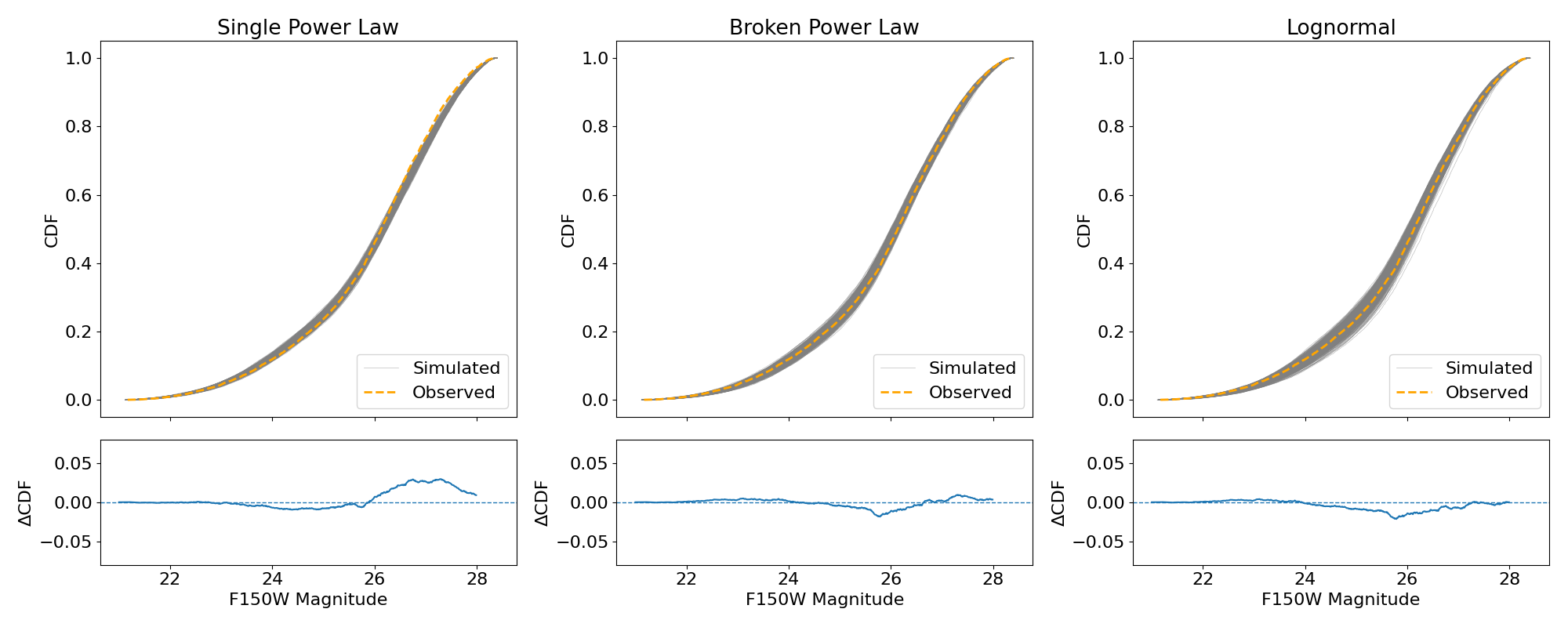}
\caption{Cumulative luminosity function of observed stars (orange dashed line) compared to posterior synthetic CMDs (gray lines). The synthetic samples represent 1000 realizations drawn from the posterior parameter distributions. The lower panels show the corresponding differences in cumulative distribution functions: $\Delta$CDF = (observed CDF) $-$ (mean of simulated CDFs).
\label{fig:lum}}
\end{figure*}

To qualitatively assess the goodness-of-fit, we draw 1000 samples from the posterior for each model and use them to simulate CMDs. From these simulations, we compute cumulative luminosity functions in the F150W band and compare them to the luminosity function of the observed CMD; the comparison is presented in Figure~\ref{fig:lum}. Overall, the simulated data based on the BPL and LN posteriors show good agreement with the observations. In contrast, the SPL model predicts a comparatively more bottom-light IMF, leading to a poorer match to the observed luminosity function.

\begin{figure*}[h]
                \centering
                 \includegraphics[width=0.9\linewidth]{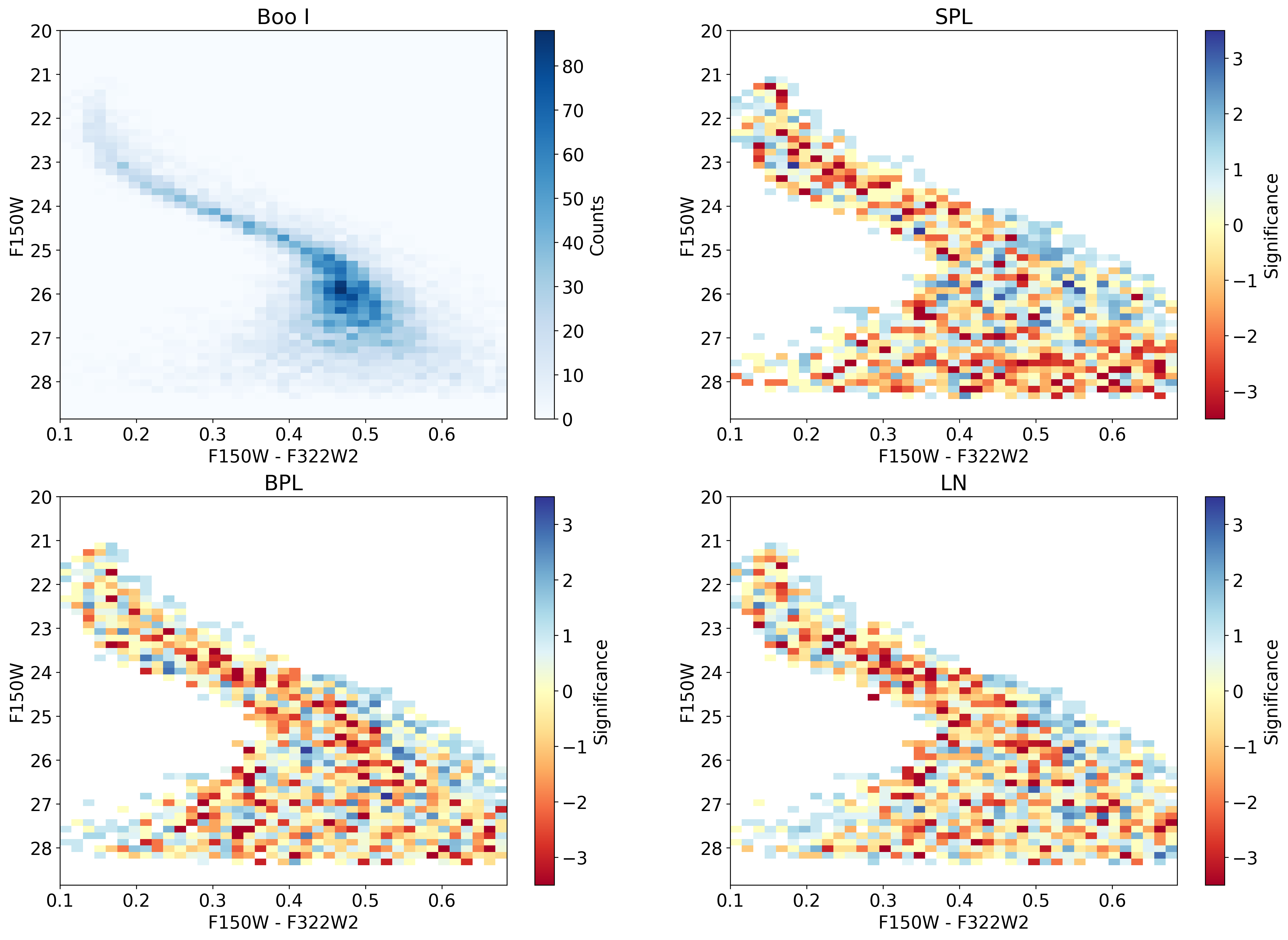}
\caption{Two-dimensional histogram of the observed CMD (top-left panel). The remaining three panels show the corresponding significance maps for each model, as defined in Equation~\ref{eq:sig}. The maps are displayed using a diverging colormap: yellow indicates good agreement, blue regions indicate bins where the observed CMD contains more stars, and red regions indicate bins where the posterior CMD contains more stars.
\label{fig:com_all}}
\end{figure*}

We also visually inspect the fit by comparing the observed and posterior CMDs. From the posterior distribution, we generate 1000 sampled CMDs and concatenate them into a single combined CMD. We then randomly draw 0.1\% of the combined entries to construct a representative CMD that captures the underlying multidimensional parameter distribution. Next, we construct two-dimensional histograms for both the observed CMD and the posterior-sampled CMD. To quantify their difference, we define the significance as
\begin{equation}
\label{eq:sig}
    S = \frac{N_{obs} - N_{pos}}{\sqrt{N_{obs}}}
\end{equation}
which represents the normalized difference in star counts between the posterior and observed CMDs in each bin, assuming Poisson noise. In Figure~\ref{fig:com_all}, the top-left panel presents the two-dimensional histogram of the observed CMD, and the remaining three panels show the corresponding significance distributions for each model. Since the adopted bin size is comparable to or smaller than the photometric uncertainty, individual bins are not expected to match exactly; however, the overall agreement between the model and observed CMDs remains reasonable.

\subsubsection{Quantitative Assessment}

To derive a more quantitative assessment of the goodness-of-fit and see whether the model predictions are consistent with our observed data, we use a goodness-of-fit criterion defined in \cite{gen18}, which is based on \cite{lem16} with the following steps:

\begin{enumerate}[label=(\roman*)]
\item Randomly draw two subsets of the posterior CMDs, a training set $\{x_j\}_{j \in J}$, and a testing set  $\{x_k\}_{k \in K}$
\item Compute, for each $k \in K$, the average distance between the CMD $x_k$ and all CMDs in the training set:
\begin{equation}
    \mathcal{D}_{\mathrm{post},k}
    =
    \frac{1}{N_j}
    \sum_{j \in J}
    \left\|
        \mu_{\mathrm{Nys}}(x_j)
        -
        \mu_{\mathrm{Nys}}(x_k)
    \right\|_2,
\end{equation}
in which $\left\|\mu_{\mathrm{Nys}}(x_j)-\mu_{\mathrm{Nys}}(x_k)\right\|_2$ is the Euclidean distance in the finite-dimensional Nystr\"{o}m feature space between two CMDs as defined in Section \ref{sec:kernel}.

\item Compute the average pairwise Euclidean distance between the observed CMD and each training CMD:

\begin{equation}
    \mathcal{D}_{\mathrm{post},obs}
    =
    \frac{1}{N_j}
    \sum_{j \in J}
    \left\|
        \mu_{\mathrm{Nys}}(x_{obs})
        -
        \mu_{\mathrm{Nys}}(x_j)
    \right\|_2.
\end{equation}
\item Compare $\mathcal{D}_{post, obs}$ with the distribution of $\mathcal{D}_{post, k}$ and see if it falls outside some pre-defined quantile.
\end{enumerate}

\begin{figure*}[h]
                \centering
                 \includegraphics[width=1\linewidth]{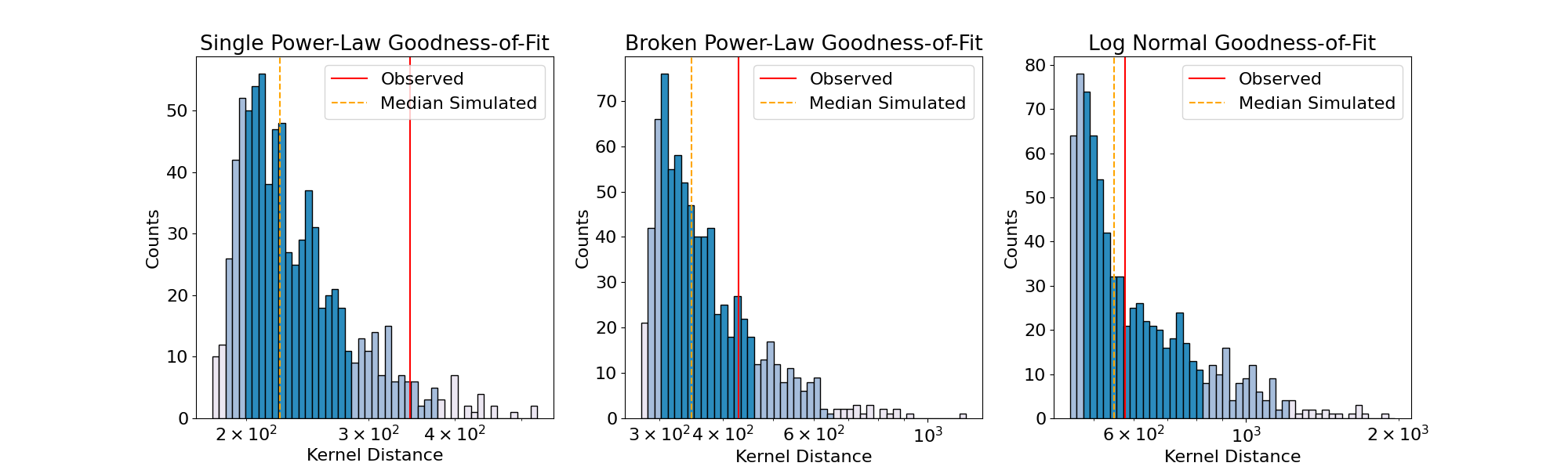}
\caption{Goodness-of-fit measurements for each IMF model. The histogram colors indicate the 68\%, 95\%, and 99\% confidence intervals of $\mathcal{D}{\mathrm{post},k}$, with the median marked by the orange dashed line. The red solid line denotes the observed value, $\mathcal{D}{\mathrm{post},\mathrm{obs}}$. The SPL model exhibits a substantially larger degree of inconsistency than the other two models.
\label{fig:gof}}
\end{figure*}

Figure~\ref{fig:gof} presents the results of the goodness-of-fit analysis for the three IMF models. The BPL and LN models show strong agreement with the observations, with consistency tighter than the 68\% confidence level, whereas the SPL model is disfavored at the 68\% confidence level. 

\subsubsection{Model-Discrimination Tests}

Beyond the apparent mismatch between the data and the SPL model, we perform an independent validation test to quantify whether our data and inference framework can distinguish between SPL and LN models, thereby assessing whether the SPL can be rejected as the underlying IMF of Boo I. We note that model discrimination depends on target- and observation-specific properties, such as completeness, the number of detected stars, and the limiting magnitude. We therefore restrict our test to the parameter space relevant to Boo I and its observational setup, without generalizing the resulting discrimination rates to the full model parameter spaces.

 Specifically, we draw 100 synthetic CMDs from the posterior distributions of the SPL and LN models, respectively. This restricts our analysis to the parameter region relevant to Boo I, rather than extending the test to the full parameter spaces of the models. For each synthetic CMD, we independently fit both the SPL and LN models using our full inference pipeline. For each fitted model, we compute the goodness-of-fit statistic defined in the previous section by comparing the observed distance metric, $D_{post, obs}$, with the posterior predictive distribution of $D_{post, k}$. Since smaller values of $D_{\rm post}$ indicate better agreement between the model and the data, the pipeline is considered to favor one model over another when $D_{\rm post,obs}$ lies at a lower quantile of the $D_{\rm post,k}$ distribution, indicating that the model provides a better description of the synthetic CMD.

  \begin{figure}[h]
                \centering
                 \includegraphics[width=1\linewidth]{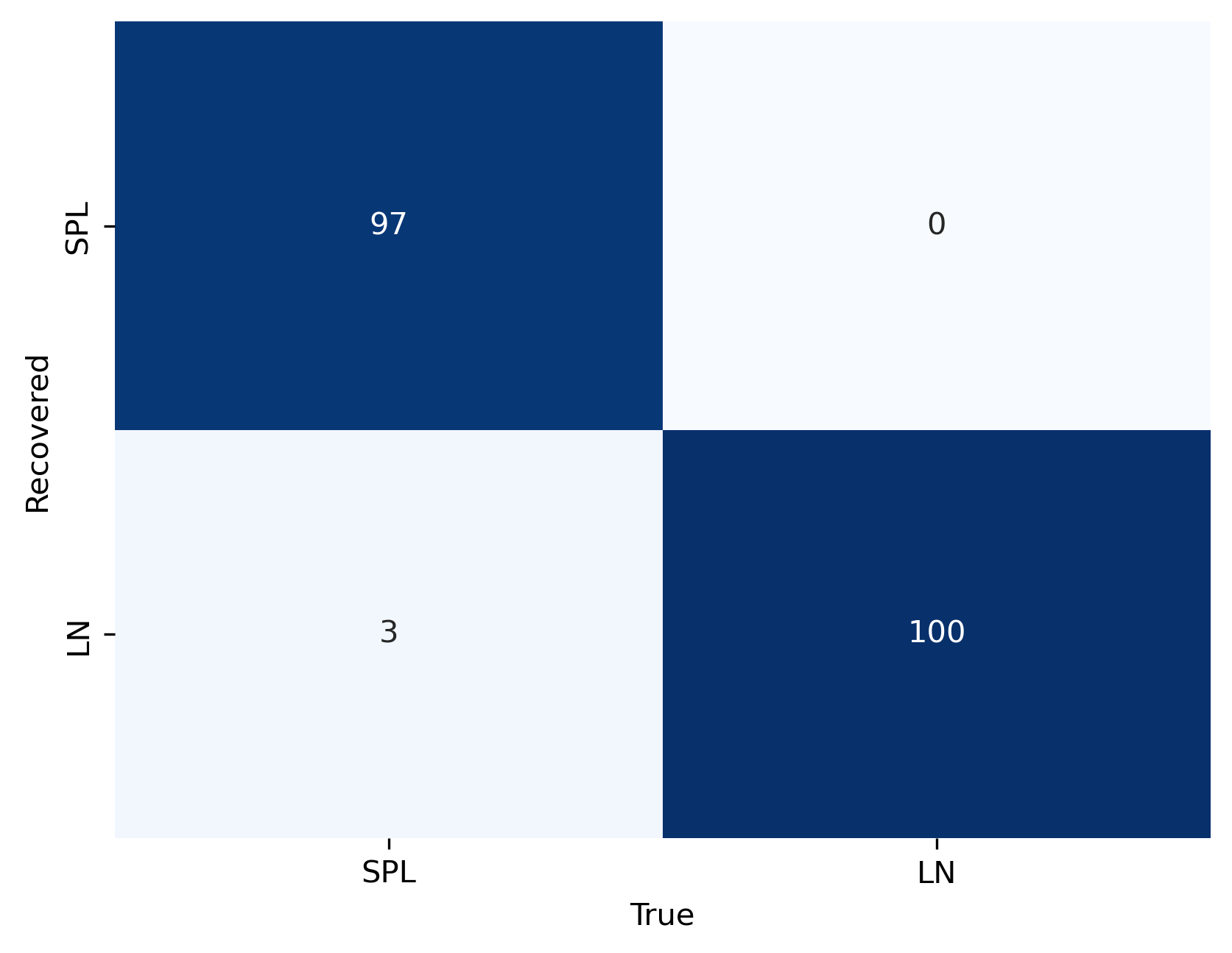}
\caption{Confusion matrix for the IMF model recovery tests. The x-axis shows the true underlying IMF model used to generate the synthetic CMDs, while the y-axis shows the IMF model preferred by the inference pipeline. Each entry represents the fraction of synthetic CMDs classified as a given model. The pipeline correctly identifies 97\% of SPL realizations and 100\% of LN realizations, demonstrating its ability to distinguish between IMFs with and without a characteristic peak mass.
\label{fig:conf}}
\end{figure}

Figure~\ref{fig:conf} shows the confusion matrix from these tests, where the x-axis indicates the true underlying IMF model and the y-axis indicates the model preferred by the pipeline. When the underlying IMF is an SPL, the pipeline correctly recovers the true model in 97\% of cases, with only 3\% of realizations being incorrectly classified as LN. When the underlying IMF is LN, the pipeline recovers the correct model in 100\% of cases. These results demonstrate that with our sample and completeness, our inference framework has sufficient discriminating power to distinguish between SPL and LN models. Combined with the poor fit of the SPL model to the observed luminosity function, these results suggest that the data favor an IMF with a characteristic peak mass, while a single power-law IMF, which diverges toward the low-mass end, is insufficient to reproduce the observations.

\section{Discussions}
\label{sec:discussions}

\subsection{Comparison to Literature Studies of Boo I IMF}

Given its proximity to the MW and its relatively high surface brightness, Boo I has been the subject of multiple studies aimed at constraining the low-mass IMF. \cite{gen18a} analyzed the IMFs of six UFDs, including Boo I, using deep HST/ACS photometry down to $\sim0.45$\msun. They obtained a best-fit single power-law slope of $\alpha = -1.84$ and best-fit log-normal parameters of $M_c = 0.33$ and $\sigma = 0.59$. Later, \citet{fil22} used HST ACS/WFC imaging to probe the IMF down to $\sim 0.3$\msun. Their results include a single power-law slope of $\alpha = -1.95^{+0.93}_{-1.04}$; broken power-law slopes of $\alpha_{\rm low} = -1.67^{+0.48}_{-0.57}$ and $\alpha{\rm high} = -2.57^{+0.93}_{-1.04}$; and log-normal parameters of $M_c = 0.17^{+0.05}_{-0.11}$\msun and $\sigma = 0.49^{+0.13}_{-0.20}$. It is worth noting that in \cite{gen18a} and our analysis, we treat binary systems as a single mass and fit the system IMF, while \cite{fil22} fitted the single-star IMF; thus, the results should not be directly compared numerically. Although previous studies did not reach the IMF peak mass and were therefore subject to large uncertainties, our deeper observations extend to the peak mass and arrive at a consistent qualitative conclusion: when IMF models with a characteristic peak mass (i.e., broken power-law or lognormal forms) are adopted, the Boo I IMF is consistent with that of the MW. In contrast, the single power-law slopes are not statistically consistent, a discrepancy that aligns with the prediction of \cite{elb17} that deeper observations can artificially favor a more bottom-light IMF in the context of the SPL model.

\subsection{Comparison to Literature Studies of Extragalactic Targets in the Local Universe}

Leveraging the deep imaging capabilities of HST, numerous studies have resolved individual extragalactic stars in the local volume to constrain the stellar IMF and reconstruct the star-formation histories of nearby galaxies (see, e.g., Section 3.1 of the review by \citealt{bas10}). However, the results vary with the mass range probed; therefore, we compare only our results with previous studies that reach mass limits of $\lesssim 0.5$\msun. 

\begin{table*}[p]
\centering
\label{tab:lit}
\renewcommand{\arraystretch}{1.1}  
\begin{tabular}{l c c c c c}
\toprule
Target & Mass Range (\msun ) & SPL & BPL & LN & Reference \\
\midrule

\multirow{3}{*}{Hercules}
 & 0.52--0.77 & $\alpha = -1.2^{+0.4}_{-0.5}$ &  & $M_c = 0.4^{+0.9}_{-0.3}$ & \citealt{geh13} \\
 & \multirow{2}{*}{0.45--0.80} & $\alpha = -0.93$  &  & $M_c=0.60$ & \multirow{2}{*}{\citealt{gen18a}} \\
  &  &  &  & $\sigma = 0.69$ &  \\
\midrule

\multirow{3}{*}{Leo IV}
 & 0.52--0.77 & $\alpha = -1.3^{+0.8}_{-0.8}$  &  & $M_c = 0.4^{+2.1}_{-0.3}$  & \citealt{geh13} \\
 & \multirow{2}{*}{0.45--0.80} & $\alpha = -0.82$  &  & $M_c = 0.62$  & \multirow{2}{*}{\citealt{gen18a}} \\
   &  &  &  & $\sigma = 0.69$ &  \\
\midrule

\multirow{5}{*}{Coma Ber}
 & \multirow{2}{*}{0.45--0.80} & $\alpha = -1.66$  &  & $M_c = 0.30$  & \multirow{2}{*}{\citealt{gen18a}} \\
 &  &  &  & $\sigma = 0.67$  &  \\
 & \multirow{3}{*}{0.20--0.75} & $\alpha = -1.45^{+0.29}_{-0.30}$  & $\alpha_{\rm low} = -1.18^{+0.49}_{-0.33}$ & $M_c = 0.33^{+0.16}_{-0.17}$ & \multirow{3}{*}{\citealt{gen18}} \\
 &  &  & $\alpha_{\rm high} = -1.88^{+0.43}_{-0.49}$ & $\sigma = 0.68^{+0.16}_{-0.12}$ &  \\ &  & & $M_b = 0.56^{+0.12}_{-0.12}$  &  \\
\midrule

 \multirow{2}{*}{CVn II} & \multirow{2}{*}{0.45--0.80} & $\alpha = -1.00$  &  & $M_c=0.56$ & \multirow{2}{*}{\citealt{gen18a}} \\
  &  &  &  & $\sigma = 0.67$ &  \\
\midrule

 \multirow{2}{*}{UMa I} & \multirow{2}{*}{0.45--0.80} & $\alpha = -1.46$  &  & $M_c=0.53$ & \multirow{2}{*}{\citealt{gen18a}} \\
  &  &  &  & $\sigma = 0.64$ &  \\
\midrule

 \multirow{3}{*}{Ret II} & \multirow{2}{*}{0.20--1.00} & $\alpha = -1.79 ^{+0.29}_{-0.27}$  & $\alpha_{\rm low} = -1.77^{+0.40}_{-0.38}$  & $M_c=0.20^{+0.06}_{-0.15}$ & \multirow{2}{*}{\citealt{fil24}} \\
  &  &  & $\alpha_{\rm high} = -2.00^{+1.37}_{-1.07}$ & $\sigma = 0.65^{+0.25}_{-0.20}$ &  \\
\midrule

 \multirow{2}{*}{UMa II} & \multirow{2}{*}{0.20--1.00} & $\alpha = -2.19 ^{+0.36}_{-0.28}$  & $\alpha_{\rm low} = -2.31^{+0.39}_{-0.36}$  & $M_c=0.15^{+0.05}_{-0.10}$ & \multirow{2}{*}{\citealt{fil24}} \\
  &  &  & $\alpha_{\rm high} = -1.60^{+1.52}_{-0.64}$ & $\sigma = 0.53^{+0.13}_{-0.26}$ &  \\
\bottomrule
\end{tabular}
\caption{Summary of IMF parameters for UFD galaxies derived from deep HST imaging. Uncertainties correspond to the 68\% confidence interval when reported. Note that \citet{geh13} fixed the lognormal $\sigma$ parameter, while \citet{fil24} fixed the break mass $M_b$ in their broken power-law fits.}
\label{tab:comparison}
\end{table*}

Table~\ref{tab:comparison} summarizes IMF parameters for UFD galaxies inferred from deep HST imaging \citep{geh13, gen18a, gen18, fil24}. We note that these studies adopted different modeling assumptions, including whether certain parameters are fixed during fitting and whether binaries are treated in terms of single-star or system masses, so the results are not directly comparable in a strictly numerical sense. However, some qualitative trends emerge from the comparison. In particular, all studies infer single power-law IMF slopes shallower than the canonical Salpeter value of $\alpha = -2.35$ \citep{sal55}. The inferred slopes also appear to correlate with observational depth, likely reflecting the increased leverage on the low-mass regime afforded by deeper data rather than an intrinsic variation alone.

 \begin{figure*}[p]
                \centering
                 \includegraphics[width=0.65\linewidth]{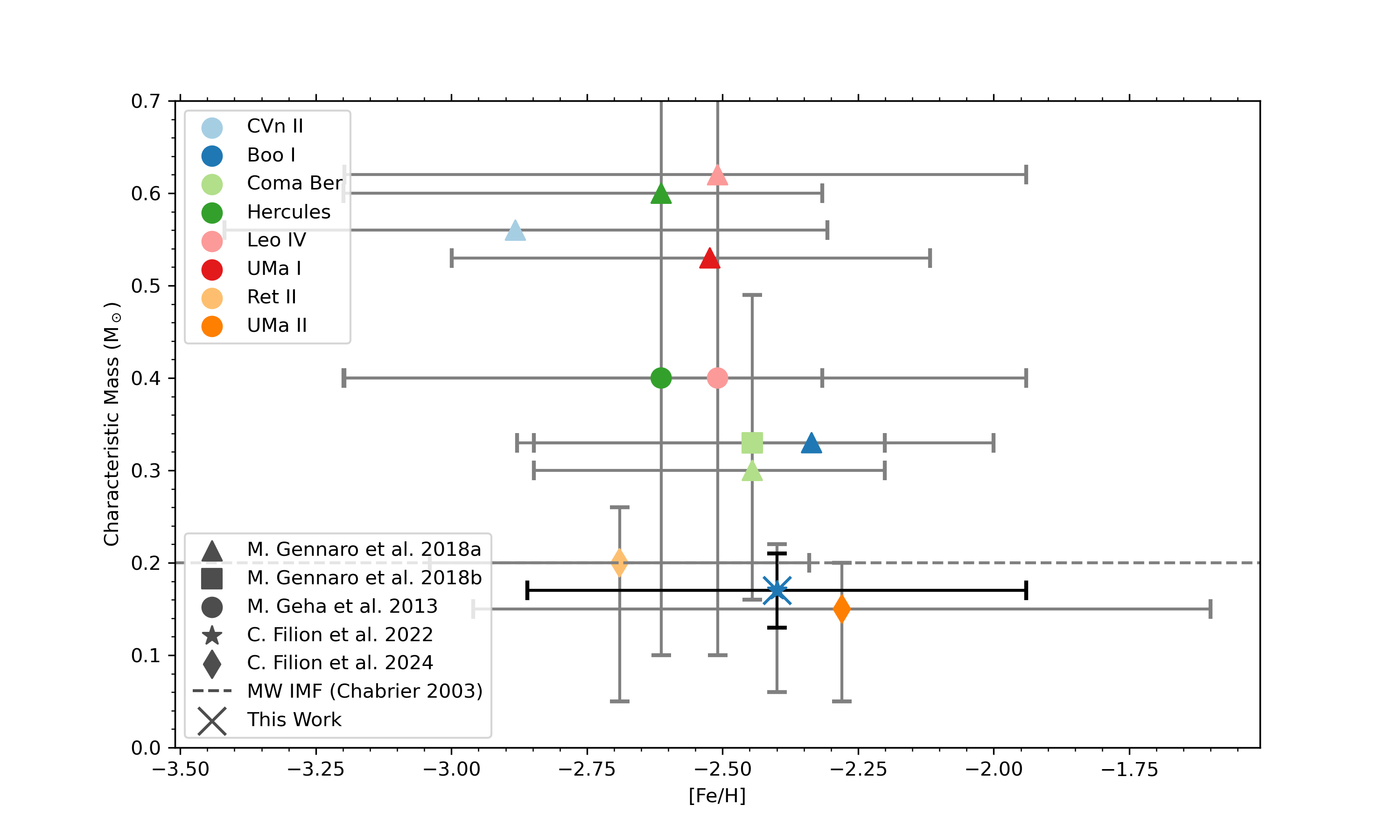}
\caption{Comparison of the best-fit lognormal characteristic mass with literature IMF values as a function of metallicity. The [Fe/H] represents the values adopted in each study and may vary for the same target. Marker shapes denote the literature sources, while colors represent the target objects. The canonical MW IMF value is shown as a dashed line. There is a mild trend for the characteristic mass to increase with lower [Fe/H], though the change is not statistically significant and correlates with the observation depth.
\label{fig:lit}}
\end{figure*}

For the lognormal model, our result is aligned with the previously suggested trend of decreasing characteristic mass $M_c$ with increasing [Fe/H], as illustrated in Figure~\ref{fig:lit}, although the trend is not statistically significant given the current uncertainties. There is the theoretical expectation that the IMF should be bottom-light (high $M_c$) in low [Fe/H] star-forming environments (e.g., \citealt{kro13} and references therein), and a similar trend is observed with JWST in the MW \citep{yas24}. In particular, \cite{sha22} proposed a transition from the primordial regime (metal-poor, gas fragmentation dominated by molecular hydrogen cooling) to the modern regime (metal-rich, dust and feedback dominated regime) at $Z\sim 10^{-2}Z_{\odot}$, accompanied by a significant shift in the characteristic mass. Given that Boo I lies on the relatively metal-rich end of currently observed UFDs \citep{bro14, geh26} with [Fe/H]=-2.4, we cannot rule out the possibility that it remains above this transition threshold and therefore retains a MW-like IMF. In contrast, the IMF of more metal-poor UFDs may exhibit a higher characteristic mass.  The mass--metallicity relation would make it difficult to obtain large sample sizes in even more metal-poor (and less massive) UFDs than Boo I.  Nonetheless, our analysis demonstrates that JWST/NIRCam observations can constrain $M_c$ with sufficient precision that, for more data points of deeply observed UFDs, it will be possible to statistically discriminate between metallicity-dependent and universal IMF scenarios in the low-metallicity regime.

\section{Conclusions}
\label{sec:conclusions}

With JWST/NIRCam, we resolved $\sim 10,000$ member stars of the Boo I UFD reaching $\lesssim$0.15\msun to probe the low-mass IMF deeper than the canonical MW IMF peak mass. We use the simulation-based inference technique to infer the single power-law, broken power-law, and lognormal IMF models through forward modeling the stellar population of Boo I and compare the synthetic and observed CMD. Our analysis shows that:

\begin{enumerate}
    \item The best-fit single power-law slope is inconsistent with a Salpeter IMF, and the model does not reproduce the observed luminosity function as well as models with a characteristic peak mass. Our goodness-of-fit measurement and model discrimination tests indicate a preference against the single power-law model, suggesting that an IMF with a characteristic peak mass provides a more suitable description of the observed stellar population in Boo I.
    \item Conditional on the presence of a break within the 0.35-0.65 \msun range, the broken power-law model can explain the luminosity function, and the IMF parameters are statistically consistent with those of the MW. However, if the allowed break mass range is expanded to the full observed mass range, the mass break is degenerate with the power-law slopes, which are poorly constrained in our fitting.
    \item Our best-fit lognormal IMF is consistent with both the canonical MW IMF and previous Boo I measurements. With our deeper photometry, we probe below the IMF peak mass and reduce the uncertainty on the characteristic mass to 11\%, providing further evidence on the invariance of the IMF.
\end{enumerate}

These results provide a window into star formation in the early universe. Under the assumption that Boo I is a local counterpart to a $\lesssim 10^5\msun$ stellar mass galaxy at $z\gtrsim6$ with metallicity down to -2.4, the inferred IMF is consistent with universality across both local and high-redshift environments. This suggests that the characteristic processes governing low-mass star formation may have been established through self-regulation (e.g., \citealt{hop11, yan23}) and remain largely insensitive to variations in the global environment, such as cosmic epoch or modest differences in chemical enrichment.

As our analysis demonstrates the resolving capability of JWST/NIRCam, future observations of UFDs could further confirm or challenge the universality of the IMF. For Boo I, the current observations are effectively near optimal, and deeper imaging is unlikely to be necessary: in the extremely low-mass regime, a narrow stellar-mass interval maps onto a wide luminosity range, so further increasing exposure time would only marginally improve the limiting mass. However, we emphasize that Boo I lies near a favorable balance between distance (and thus effective depth), sky coverage, and available stellar counts. As a result, extending this approach to other UFDs will likely require different observational trade-offs, such as longer exposures for more distant systems (e.g., Hercules), or wider sky coverage for nearer systems (e.g., Coma Berenices).

\section{acknowledgments}
We thank the referee for their constructive feedback.

This work is based [in part] on observations made with the NASA/ESA/CSA James Webb Space Telescope. The data were obtained from the Mikulski Archive for Space Telescopes at the Space Telescope Science Institute, which is operated by the Association of Universities for Research in Astronomy, Inc., under NASA contract NAS 5-03127 for JWST. These observations are associated with program PID3849. Some/all of the data presented in this article were obtained from the Mikulski Archive for Space Telescopes (MAST) at the Space Telescope Science Institute. The specific observations analyzed can be accessed via \dataset[doi:10.17909/05ft-v610]{https://doi.org/10.17909/05ft-v610}.

Support for program PID3849 was provided by NASA through a grant from the Space Telescope Science Institute, which is operated by the Association of Universities for Research in Astronomy, Inc., under NASA contract NAS 5-03127.

\software{\texttt{NumPy} \citep{numpy}, \texttt{SciPy} \citep{scipy}, \texttt{astropy} \citep{astropy_2013, astropy_2018, astropy_2022}, \texttt{Matplotlib} \citep{matplotlib}, \texttt{sbi} \citep{sbi}, \texttt{DOLPHOT}, \citep{dol00, wei24}}

\facility{JWST}

\clearpage
\appendix

\section{Empirical Photometric Quality Cuts}
\label{app:cuts}

\begin{figure*}[h]
                \centering
                 \includegraphics[width=0.8\linewidth]{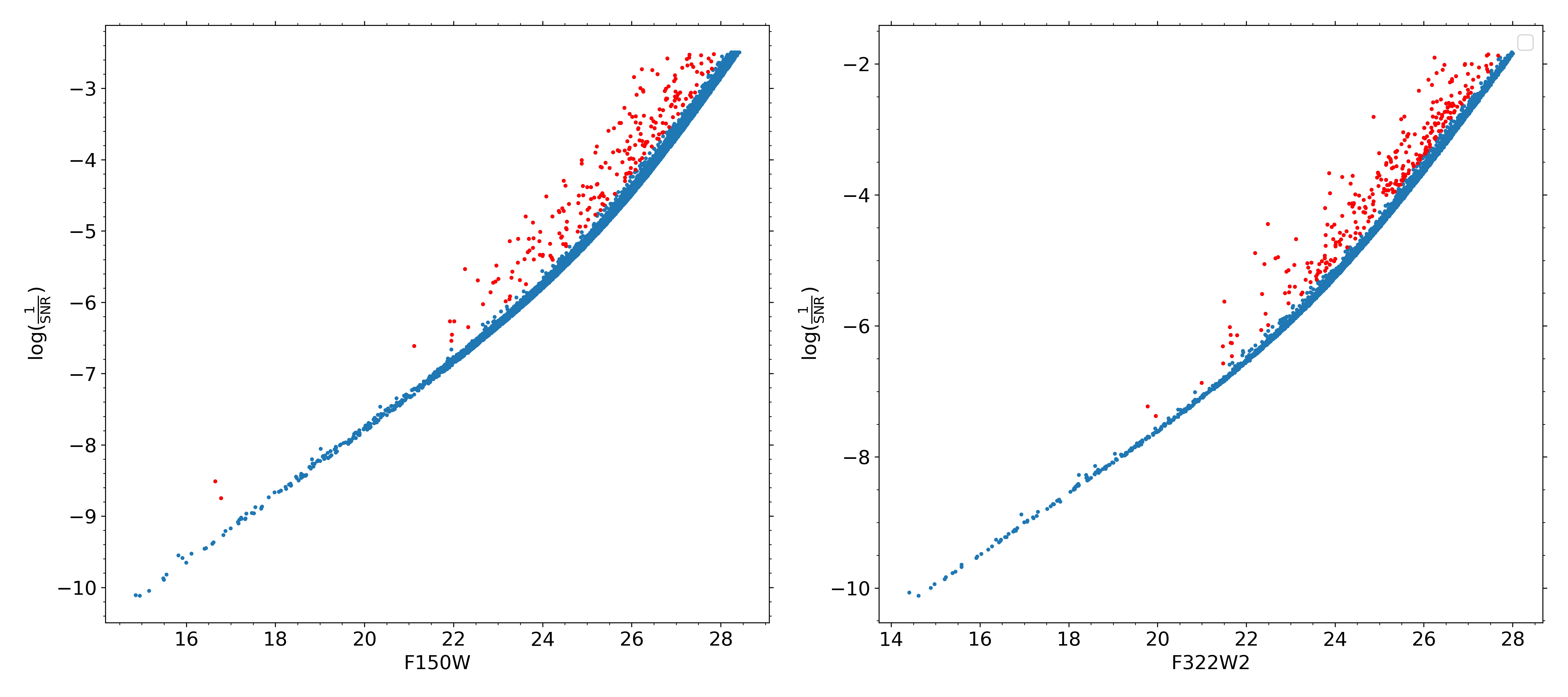}
\caption{Relation between magnitude and $\log\left(1/\mathrm{SNR} \right)$ in the F150W (left) and F322W2 (right) bands. The blue points show sources that pass the empirical SNR-based quality cut, while red points indicate rejected sources.
\label{fig:log_cut}}
\end{figure*}

In addition to the fixed photometric quality cuts described in Section~\ref{sec:qual}, we apply an empirical data-quality criterion based on magnitude and SNR. For each band, we fit an empirical relation between ${\tt \log(1/\mathrm{SNR})}$ and magnitude, and reject sources whose deviation from this relation exceeds $0.2$, as shown in Figure~\ref{fig:log_cut}. This criterion removes extracted sources with photometric uncertainties significantly larger than the typical uncertainty at a given magnitude.

\section{Artificial Star Tests}
\label{app:ast}

\begin{figure*}[h]
                \centering
                 \includegraphics[width=\linewidth]{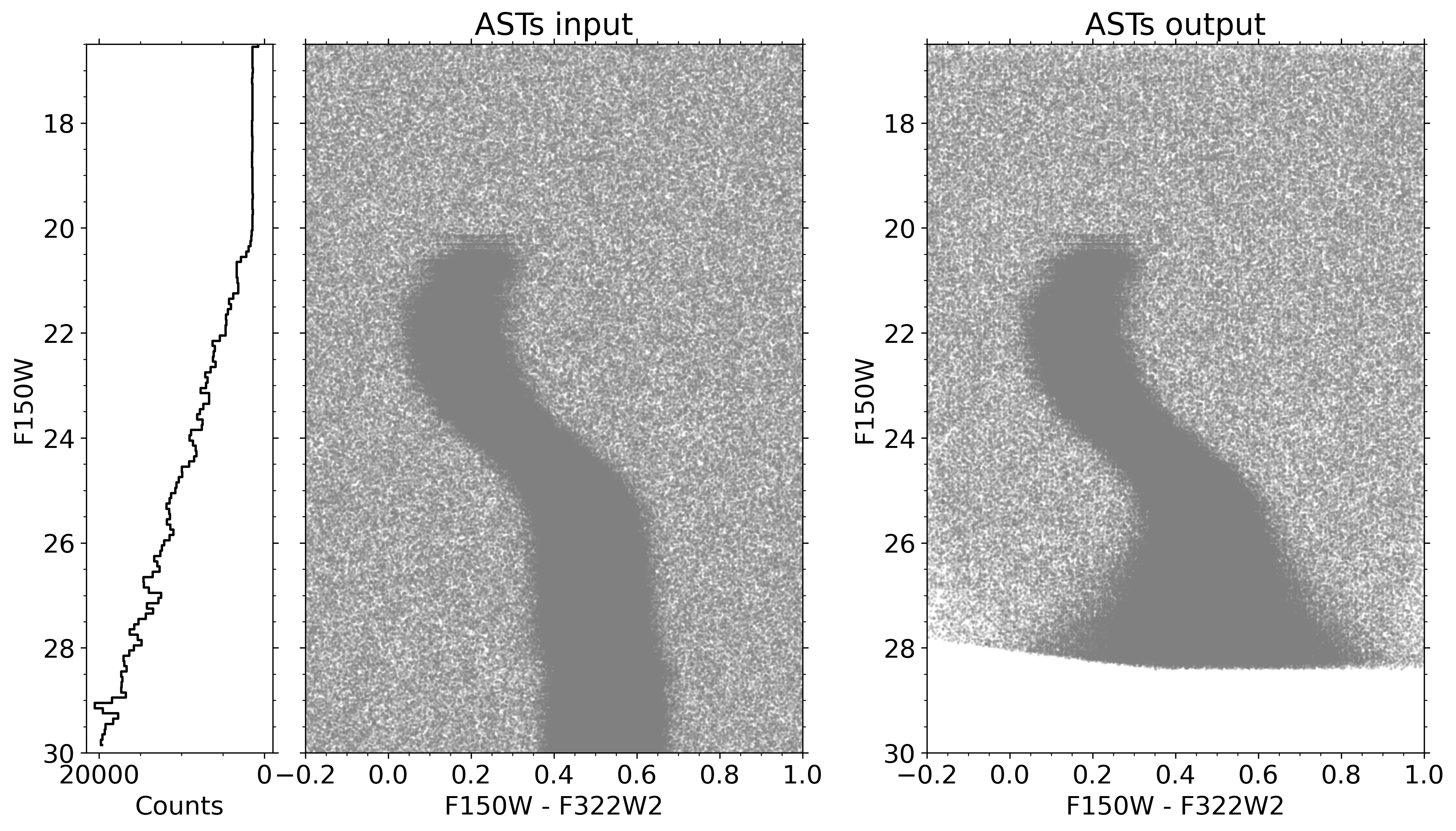}
\caption{Left: the input luminosity function of the artificial stars; Middle: the input CMD used for ASTs; Right: the recovered output CMD from ASTs after applying photometric quality cuts
\label{fig:ast_comp}}
\end{figure*}

As detailed in Section~\ref{sec:ast}, the AST input catalog consists of $10^6$ sources sampled around the Boo I main sequence, together with $2\times10^5$ sources uniformly distributed across the CMD. The left and middle panels of Figure~\ref{fig:ast_comp} show the luminosity function and CMD of the input catalog, respectively, while the right panel shows the recovered CMD after applying the same photometric quality cuts used for the science catalog. We find that recovery drops sharply beyond $F150W \gtrsim 28$, and that the lower main sequence ($F150W \gtrsim 26$) is increasingly broadened by photometric uncertainties.

\section{Spectroscopic Metallicity Distribution Function (MDF) of Boo I}
\label{app:mdf}

\begin{figure}[h]
                \centering
                 \includegraphics[width=0.6\linewidth]{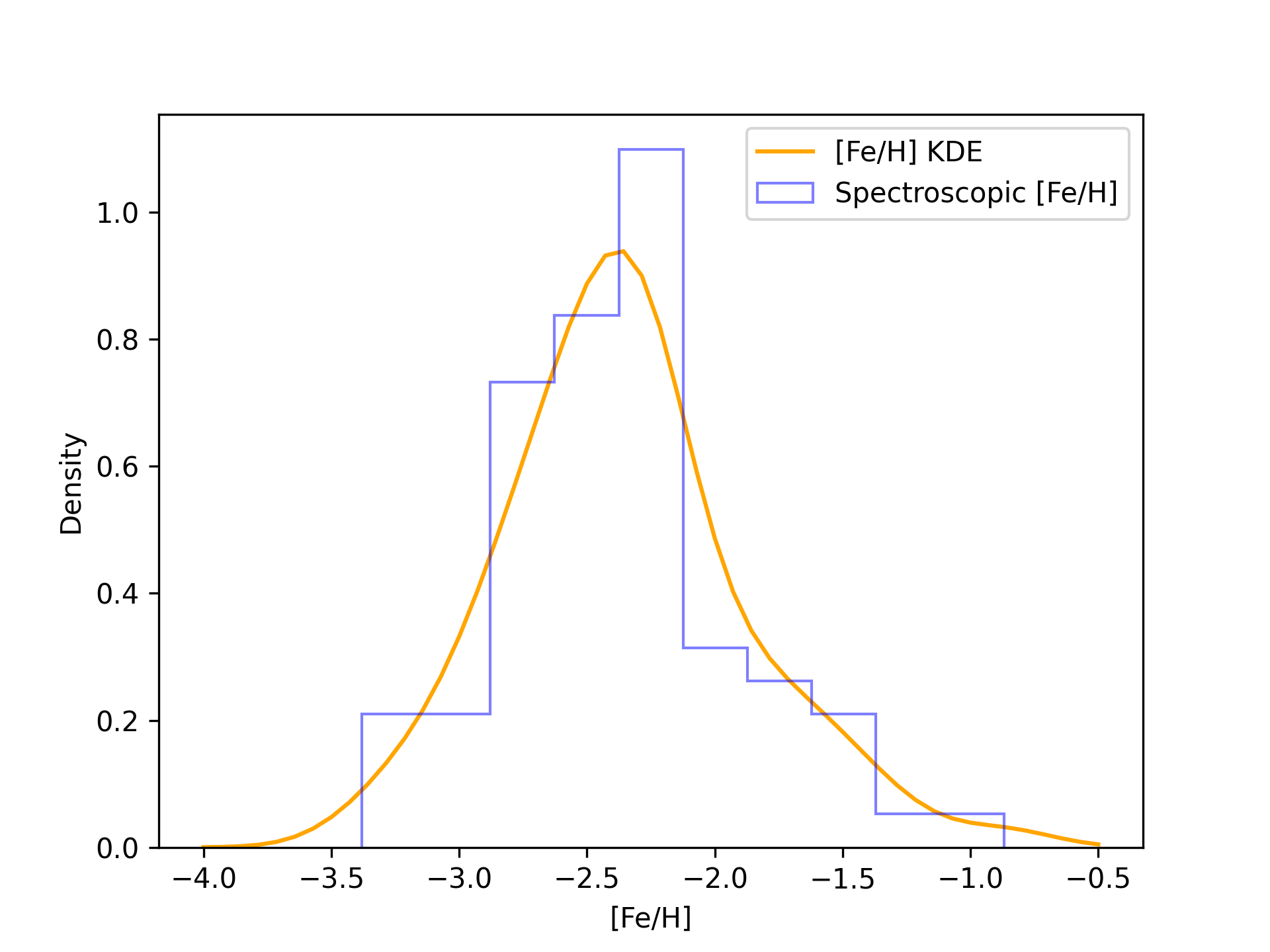}
\caption{The spectroscopic MDF of Boo I, with the blue line showing the [Fe/H] histogram and the orange curve showing the smooth probability density function by smoothing the histogram with a Gaussian Kernel Density Estimation.
\label{fig:mdf}}
\end{figure}

The metallicity distribution function is estimated based on the Keck/DEIMOS spectroscopic [Fe/H] derived in \cite{geh26}. The spectra were reduced with the \texttt{PypeIt} pipeline, and the stellar [Fe/H] was determined using the equivalent widths (EWs) of the Ca II triplet (CaT) lines at (8498.0, 8542.1, and 8662.1 {\AA}). \cite{geh26} also provided the membership probabilities $P_{mem}$ based on a set of criteria, including metallicity, proper motion, and velocities, and we selected the likely spectroscopic members of Boo I following $P_{mem}>0.5$. To mitigate the effects of binning in the discrete spectroscopic [Fe/H] histogram, we apply the Gaussian Kernel Density Estimation method to interpolate between bins and construct a smooth MDF; the resulting distribution is presented in Figure~\ref{fig:mdf}. The resulting MDF is well described by a bell-shaped distribution, peaking at [Fe/H] = -2.4 with a standard deviation of 0.46.



\section{SBI Hyperparameter Determination and Convergence Tests}
\label{app:param}

As discussed in Section~\ref{sec:kernel}, the SBI includes multiple hyperparameters, such as the number of rounds, the simulations per round, the kernel width, and the dimension for Nystr\"{o}m approximation. Ideally, the number of simulations per round should be as large as possible, but an extremely large number of simulations is unrealistic given that the simulator is hard to parallelize; therefore, we limit our choices to 5 rounds and 3000 simulations per round.

With fixed numbers of rounds and simulations, we use the procedure described in Section~\ref{sec:kernel} to empirically determine the optimal combination of Gaussian kernel width $\gamma$ and embedding dimension $n$. We first simulate reference CMDs following the physical parameters of Boo I and the canonical MW IMF parameters, incorporating photometric noise according to our ASTs. We then construct a grid of $\gamma$ values ranging from 0.5 to 50,000 in steps of $5 \times 10^{k}$, and explore embedding dimensions $n$ from 25 to 400. Then we feed the reference CMDs back to \texttt{Starwave} with each combination of $\gamma$ and $n$, testing whether the method can successfully recover the known input parameters. 

\begin{figure*}[ht]
\centering

\begin{subfigure}{0.48\textwidth}
    \centering
    \includegraphics[width=\linewidth]{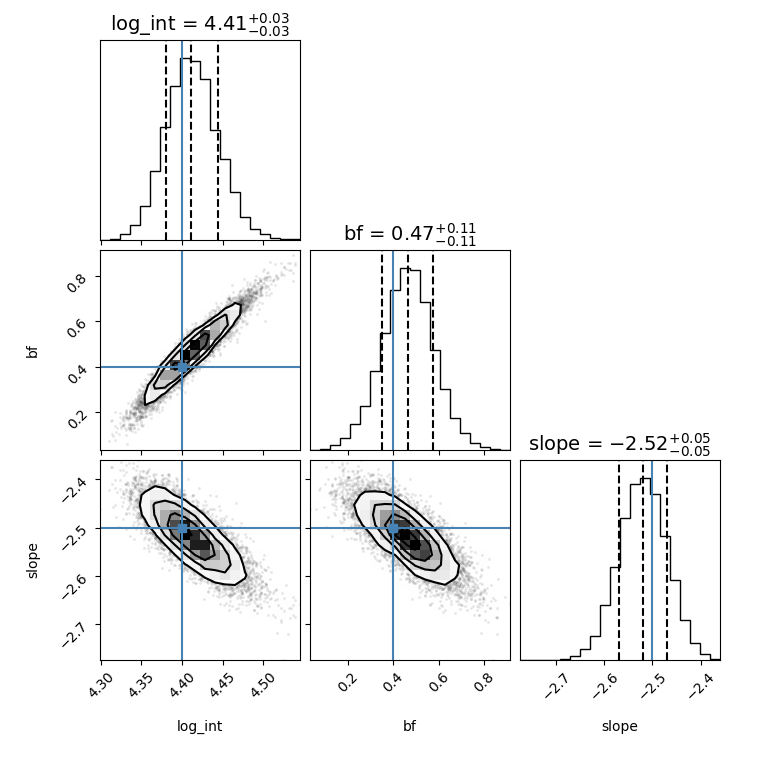}
    \caption{Single power-law}
    \label{fig:spl_conv}
\end{subfigure}\hfill
\begin{subfigure}{0.48\textwidth}
    \centering
    \includegraphics[width=\linewidth]{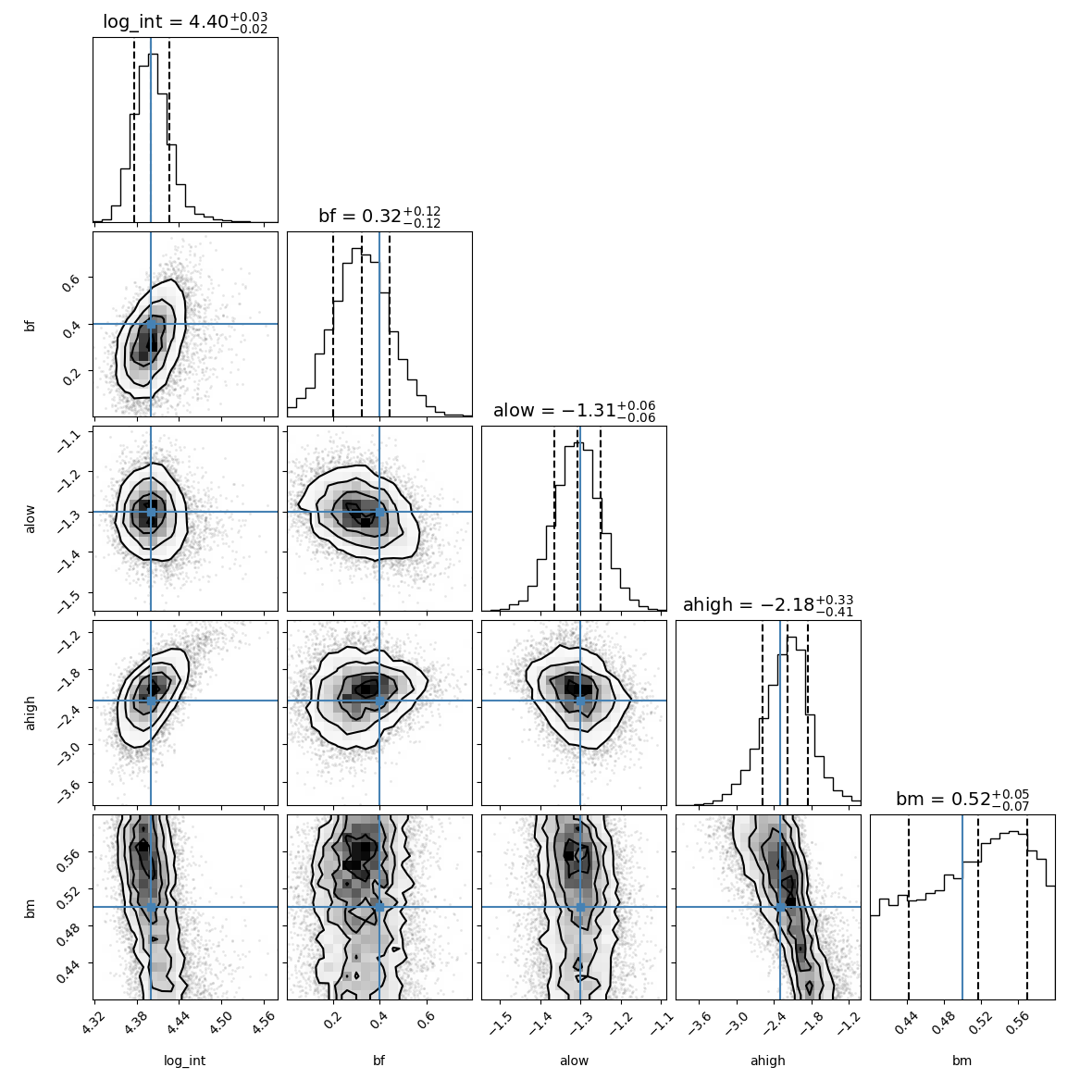}
    \caption{Broken power-law}
    \label{fig:bpl_conv}
\end{subfigure}

\vspace{0.5cm}

\begin{subfigure}{0.48\textwidth}
    \centering
    \includegraphics[width=\linewidth]{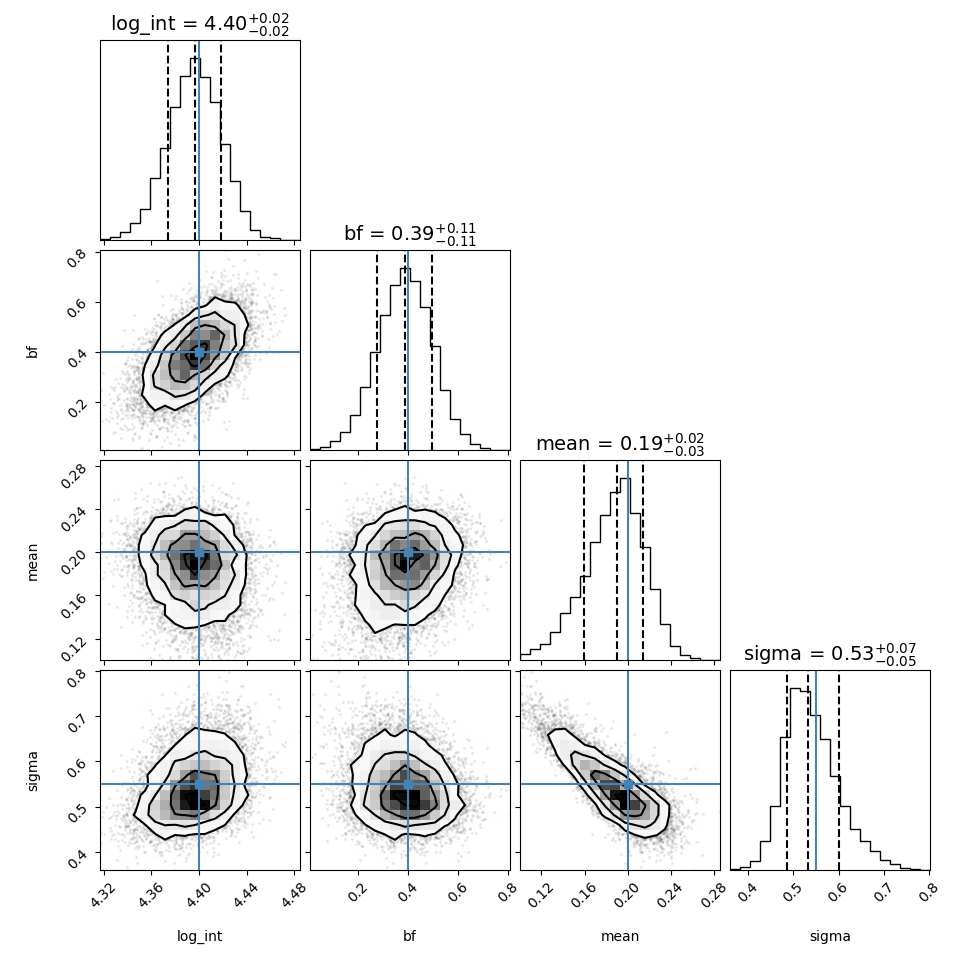}
    \caption{Log-normal}
    \label{fig:ln_conv}
\end{subfigure}
\caption{Posterior corner plots from the convergence test, with blue lines marking the true (input) parameter values used to generate the reference CMDs.}
\label{fig:conv}

\end{figure*}

The optimal hyperparameter combination is chosen according to two criteria: (1) the parameter uncertainties are minimized while the true values fall within the 68\% confidence interval, and (2) the posterior CMDs visually reproduce the luminosity function of the input CMD. Our experiments indicate that $\gamma = 50$ and $n = 50$ yield the best overall performance, and the uncertainties in the posteriors can be interpreted as intrinsic uncertainties that cannot be further reduced given our method and data, as demonstrated in Figure~\ref{fig:conv}.

\section{BPL Break Mass Priors}
\label{app:bm}

As mentioned in Section~\ref{sec:results}, when the break mass in the BPL model is allowed to vary across a range comparable to the mass range of the data, the sampler can push the break mass to the edge of the prior, effectively reducing the BPL to a SPL with another power-law slope unconstrained. To demonstrate such an effect, we generate a synthetic CMD with known parameters, and use the pipeline to recover the input parameters while allowing the break mass to change between 0.2 and 0.8 \msun.

 \begin{figure}[h]
                \centering
                 \includegraphics[width=0.6\linewidth]{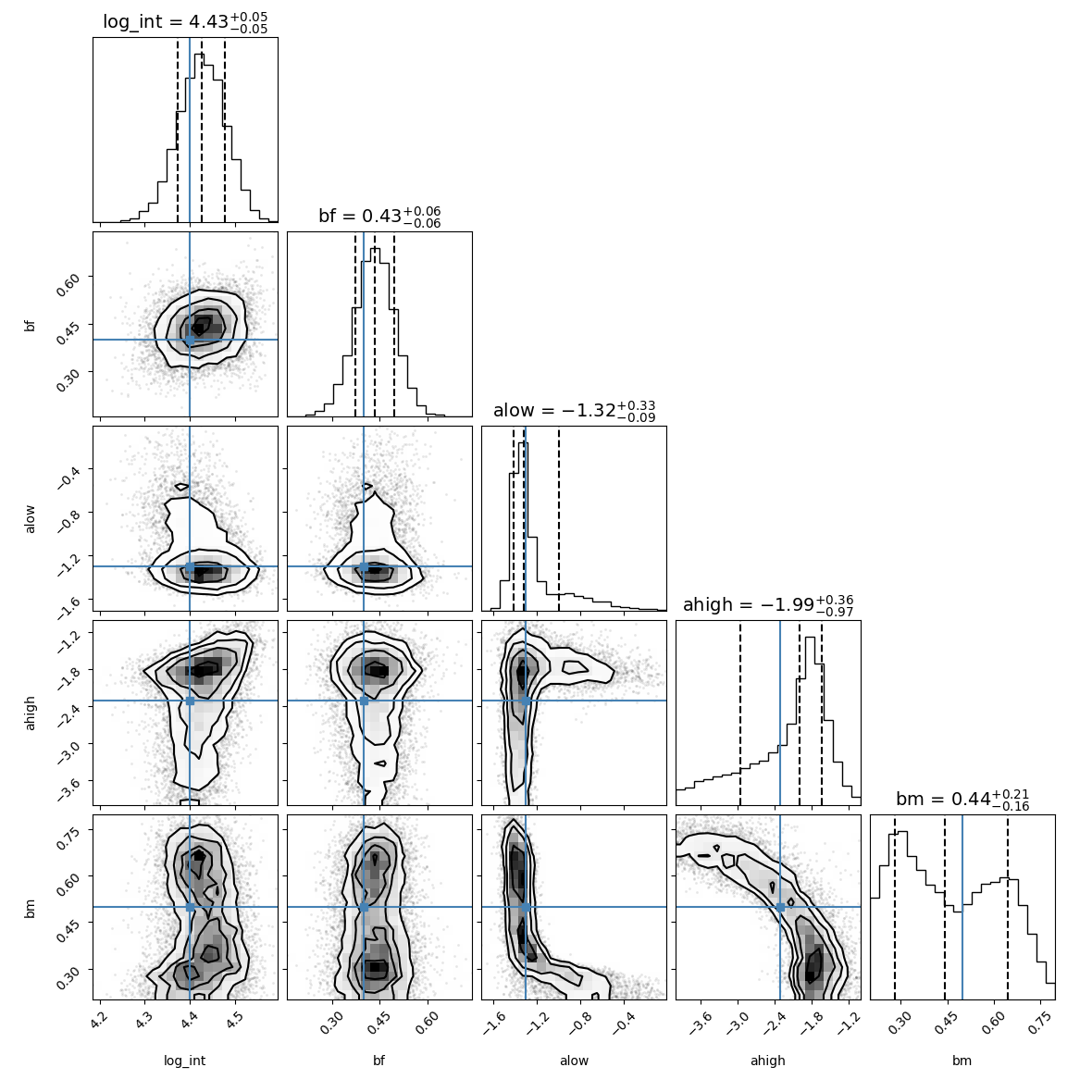}
\caption{Posterior corner plots from the BPL convergence test, with blue lines marking the true (input) parameter values used to generate the reference CMDs. The break mass is allowed to vary between 0.2 and 0.8 \msun.
\label{fig:bpl_large_bm}}
\end{figure}

The result of the convergence test is shown in Figure~\ref{fig:bpl_large_bm}. The marginal posterior distribution of the break mass is bimodal, with both peaks near the boundaries of the allowed prior range. We find that the break mass is strongly degenerate with the two power-law slopes, such that when the break mass approaches one boundary, only one of the two slopes can be effectively constrained. This demonstrates that adopting a broad break mass prior introduces strong parameter degeneracies in our inference framework given our data and completeness. Therefore, when fitting the observed data with a BPL model, we adopt a narrower break mass prior of 0.35-0.65 \msun to mitigate this degeneracy.

\section{Effects of the Stellar Population Parameters}
\label{app:phy}

To investigate the effect of the physical parameter assumption on the IMF parameters posteriors, to perform another inference in which we allow the stellar population parameters, including binary fraction (bf), distance modulus (dm), extinction (av), and age, to vary, and simultaneously fit these stellar population parameters together with the IMF parameters. We adopted flat priors for the population parameters, with ranges of 18.7-19.3 for the distance modulus (corresponding to 55-72.5 kpc), $\pm10\%$ around the fiducial value of $A_V$, and 13-14 Gyr for the stellar age. It is worth noting here that we keep the [Fe/H] fixed, since we use an empirical MDF distribution that is not parametrized.

\begin{figure*}[h]
\centering

\begin{subfigure}{0.48\textwidth}
    \centering
    \includegraphics[width=\linewidth]{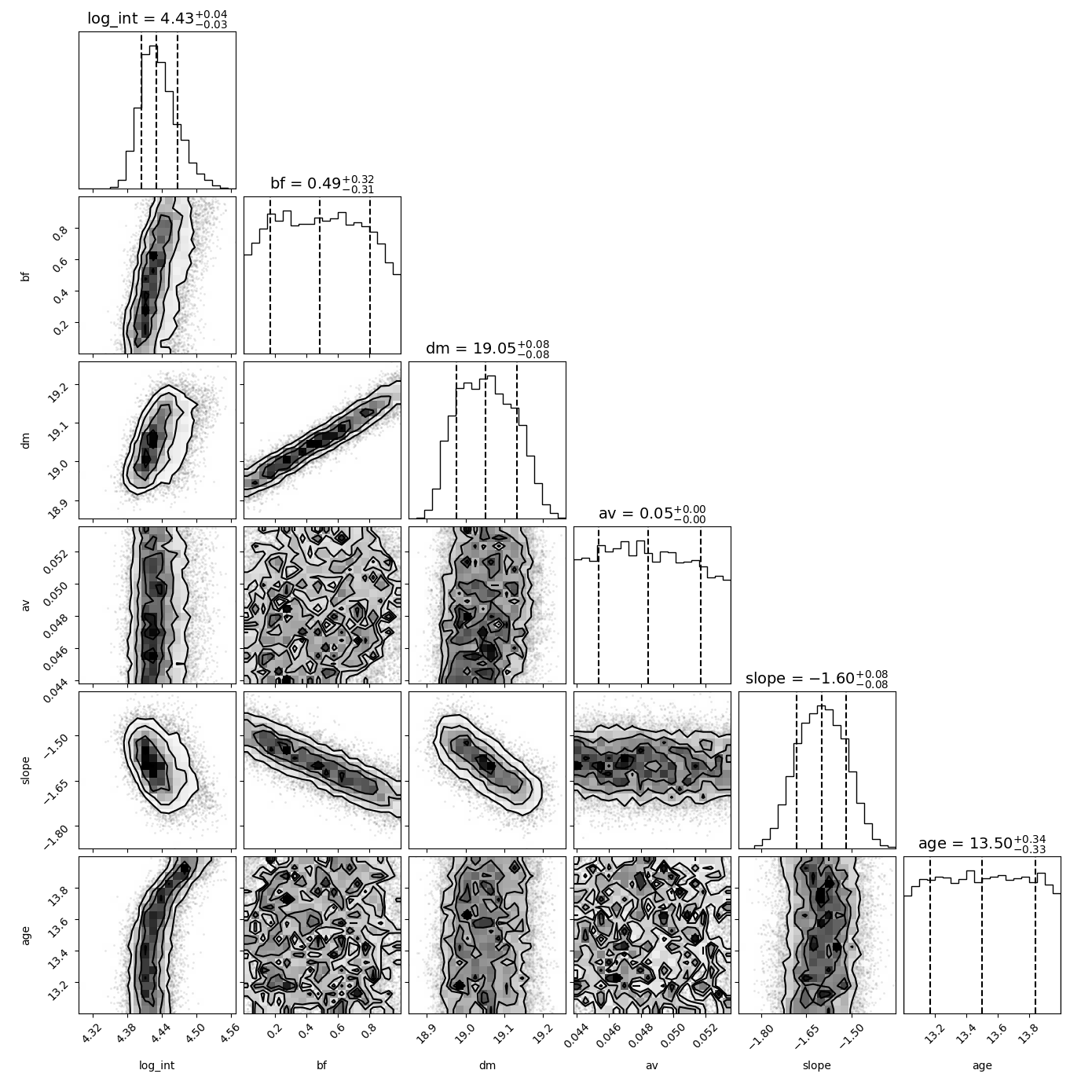}
    \caption{Single power-law}
    \label{fig:spl_phy}
\end{subfigure}\hfill
\begin{subfigure}{0.48\textwidth}
    \centering
    \includegraphics[width=\linewidth]{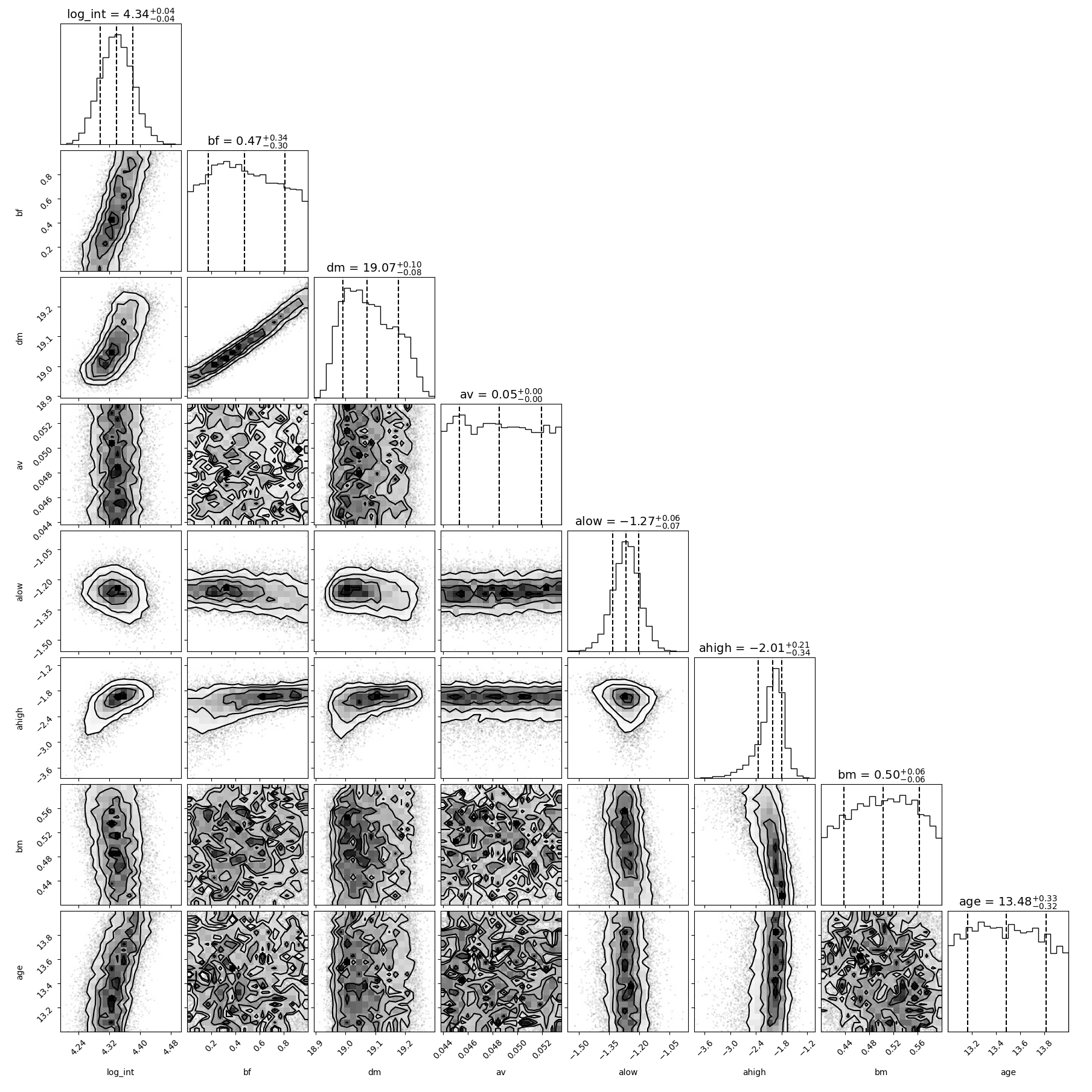}
    \caption{Broken power-law}
    \label{fig:bpl_phy}
\end{subfigure}

\vspace{0.5cm}

\begin{subfigure}{0.48\textwidth}
    \centering
    \includegraphics[width=\linewidth]{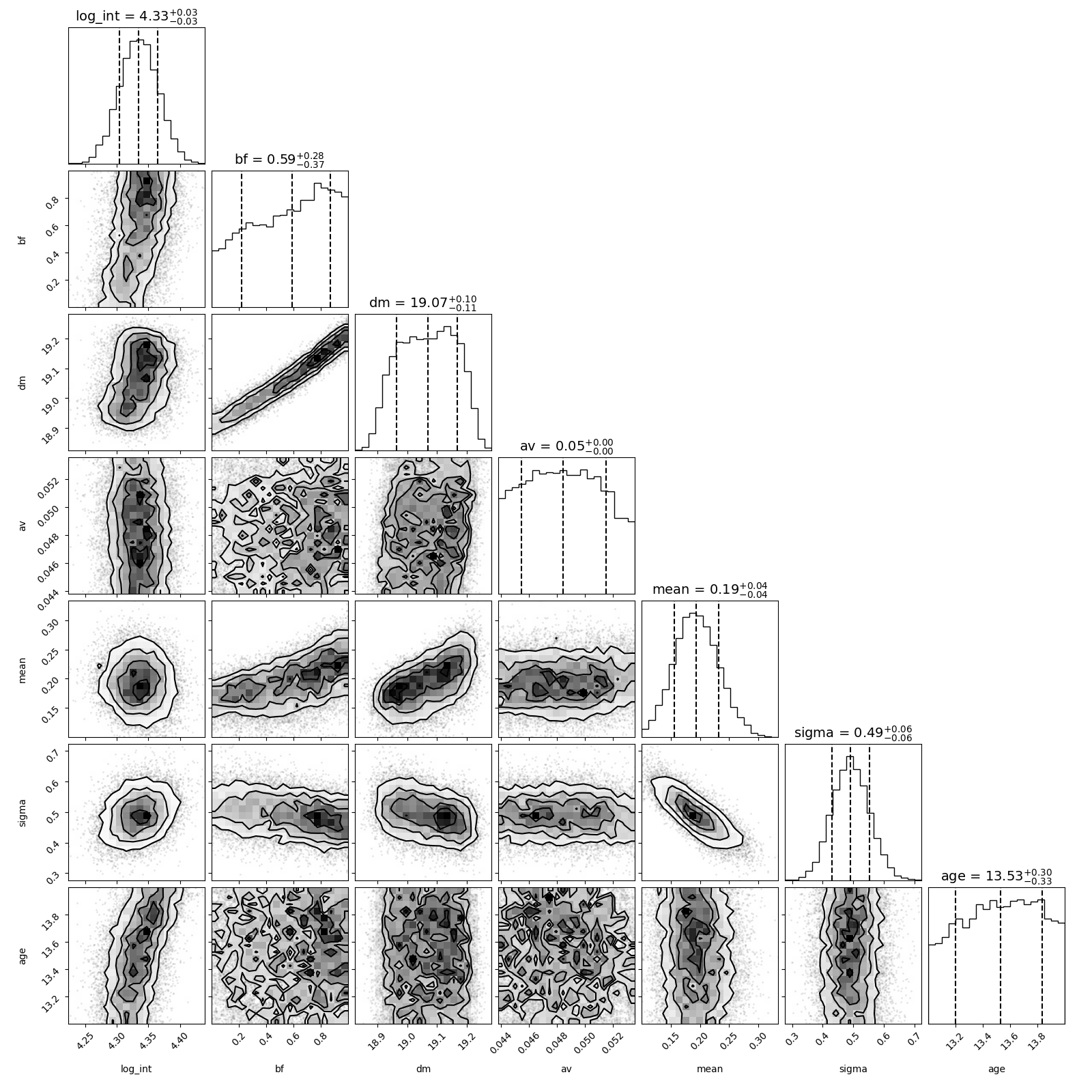}
    \caption{Log-normal}
    \label{fig:ln_phy}
\end{subfigure}\hfill
\begin{subfigure}{0.48\textwidth}
    \centering
    \includegraphics[width=\linewidth]{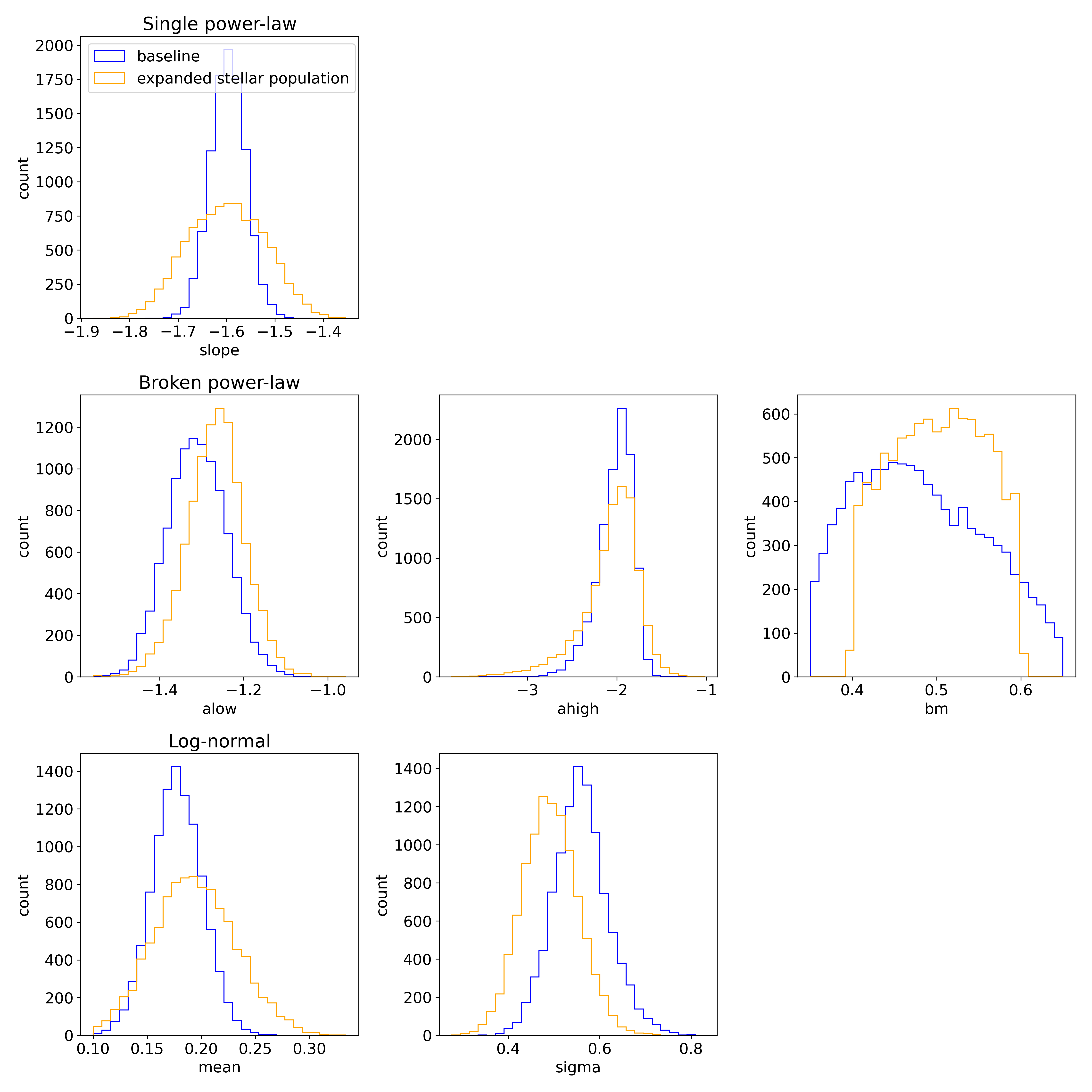}
    \caption{Marginal distributions comparison}
    \label{fig:compare}
\end{subfigure}

\caption{Corner plots of the posterior distributions for the three IMF models, where the IMF and stellar population parameters are fitted simultaneously. Vertical dashed lines mark the 16th, 50th, and 84th percentiles of the marginal distributions. Comparisons of the marginalized distributions of the IMF parameters are shown in the bottom right panel, with the blue histogram showing the posterior from the baseline fit and the orange histogram from the expanded stellar population fit.}
\label{fig:phy}
\end{figure*}

The posterior corner plots are shown in Figure~\ref{fig:phy} and a comparison between the baseline and expanded stellar population models is presented in Table~\ref{tab:imf_robustness}. Because our analysis focuses on the nearly vertical part of the main-sequence, the CMD provides only a weak constraint on the binary fraction. In particular, the binary fraction is highly degenerate with the distance modulus, as a higher binary fraction or a lower distance modulus both effectively shift stars to brighter magnitudes on the CMD. The impact of extinction is minimal, as infrared magnitudes are only weakly affected. Moreover, age does not exhibit significant degeneracy with the IMF parameters, since our analysis probes stellar masses below the main-sequence turnoff. Consequently, the assumed stellar population parameters show little correlation with the IMF parameters, and our main conclusions remain robust under modest variations in the adopted physical parameters of Boo I.

\begin{table*}
\centering
\begin{tabular}{llccc}
\hline
\hline
IMF Model & Parameter & Baseline & Expanded Stellar Population Model & $\Delta/\sigma_{\rm base}$ \\
\hline

\multirow{1}{*}{Single Power Law}
& $\alpha$ &
$-1.60^{+0.04}_{-0.04}$ &
$-1.60^{+0.08}_{-0.08}$ &
$0.00$ \\

\hline

\multirow{3}{*}{Broken Power Law}
& $\alpha_{\rm low}$ &
$-1.31^{+0.07}_{-0.07}$ &
$-1.27^{+0.06}_{-0.07}$ &
$0.57$ \\

& $\alpha_{\rm high}$ &
$-1.99^{+0.16}_{-0.22}$ &
$-2.01^{+0.21}_{-0.34}$ &
$-0.11$ \\

& $M_{\rm break}\;(M_\odot)$ &
$0.47^{+0.09}_{-0.07}$ &
$0.50^{+0.06}_{-0.06}$ &
$0.38$ \\

\hline

\multirow{2}{*}{Lognormal}
& $M_{\rm c}\;(M_\odot)$ &
$0.18^{+0.02}_{-0.02}$ &
$0.19^{+0.04}_{-0.04}$ &
$0.50$ \\

& $\sigma$ &
$0.56^{+0.06}_{-0.06}$ &
$0.49^{+0.06}_{-0.06}$ &
$-1.17$ \\

\hline
\end{tabular}
\caption{Comparison of the inferred IMF parameters between the baseline analysis and the expanded stellar population model, where the distance modulus, extinction, age, and binary fraction are jointly varied. Quoted uncertainties correspond to the 68\% credible intervals. The final column shows the shift in the posterior median relative to the baseline uncertainty, $\Delta/\sigma_{\rm base} = (\theta_{\rm expanded}-\theta_{\rm baseline})/\sigma_{\rm base}$.
\label{tab:imf_robustness}}
\end{table*}

\clearpage
\bibliography{sample701}{}
\bibliographystyle{aasjournalv7}


\end{CJK*}
\end{document}